\documentclass[a4paper,11pt]{article}
\pdfoutput=1 
\usepackage{jheppub} 
\usepackage[T1]{fontenc} 
\usepackage{subfigure}
\usepackage{hyperref}
\usepackage{slashed}
\usepackage{cancel}
\usepackage{silence}
\usepackage{ulem}
\newcommand{\eq}[1]{\begin{align}#1\end{align}}

\title{Title}
\author[ ]{M. A. Arroyo-Ure\~na${}^{1}$,}
\author[]{G.~
Hern\' andez-Tom\'e${}^{1,2}$,}
\author[]{G.~
L\'opez-Castro${}^{1}$,}
\author[]{P.~
Roig${}^{1}$,}
\author[] {I.~
Rosell${}^{3}$.}
\affiliation{${}^1$ Departamento de F\'\i sica, Centro de Investigaci\' on y de Estudios Avanzados del Instituto Polit\' ecnico Nacional, Apartado Postal 14-740, 07000 Ciudad de M\'exico, M\' exico }
\affiliation{${}^2$ Instituto de F\'isica, Universidad Nacional Aut\'onoma de M\'exico, AP 20-364, Ciudad de M\'exico 01000, M\'exico.}
\affiliation{${}^3$  Departamento de Matem\'aticas, F\'\i sica y Ciencias Tecnol\' ogicas, Universidad Cardenal Herrera-CEU, CEU Universities, 46115 Alfara del Patriarca, Val\`encia, Spain}

\emailAdd{marco.arroyo@cinvestav.mx}
\emailAdd{ghernandez@fisica.unam.mx}
\emailAdd{glopez@fis.cinvestav.mx}
\emailAdd{proig@fis.cinvestav.mx}
\emailAdd{rosell@uchceu.es}

\begin{document} 
\title{One-loop determination of $\tau \to \pi (K)  \nu_{\tau}[\gamma]$ branching ratios and new physics tests }
\abstract{We calculate the ratios $R_{\tau / P}\equiv \Gamma \left( \tau \to P \nu_{\tau}[\gamma] \right) / \Gamma \left(P\to \mu \nu_{\mu}[\gamma]\right)$ ($P=\pi,K$) at one loop following a large-$N_C$ expansion where Chiral Perturbation Theory is enlarged by including the lightest resonances and respecting the short-distance behavior dictated by QCD. We find $\delta R_{\tau/\pi}=(0.18\pm 0.57 )\%$  and $\delta R_{\tau/K}=(0.97\pm 0.58 )\%$, where the uncertainties are induced fundamentally by the counterterms. We test the lepton universality, obtaining $\left|g_\tau/g_\mu\right|_\pi=0.9964\pm 0.0038$ and $\left|g_\tau/g_\mu\right|_K=0.9857\pm 0.0078$, and analyze the CKM unitarity, getting results at $2.1\sigma$ and $1.5\sigma$ from unitarity via $|V_{us}/V_{ud}|$ and $|V_{us}|$, respectively. We also update the search for non-standard interactions in $\tau$ decays. As a by-product, we report the theoretical radiative corrections to the $\tau \to P \nu_{\tau}[\gamma]$ decay rates: $\delta_{\tau \pi} = -(0.24 \pm 0.56) \%$ and $\delta_{\tau K} = -(0.15 \pm 0.57) \%$.}
\maketitle
\flushbottom
\section{Introduction}
A fundamental result of the Standard Model~\cite{SM} is that the three lepton families couple to the electroweak gauge bosons with the same intensity $g_{\ell}$ ($\ell=e,\, \mu,\, \tau)$. This property, known as lepton universality (LU), has been tested in a large diversity of weak interaction processes and most of the results are compatible with LU.  However, a few anomalies observed in semileptonic $B$ meson decays~\cite{Albrecht:2021tul} seem to indicate a persistent deviation from the expected predictions. So far, such anomalies are not statistically significant enough to claim evidence of new physics (NP), but these intriguing data have drawn the interest of the particle physics community suggesting several possible explanations of NP. Then, one of the most active tasks in both the theoretical and experimental particle physics program is the study and review of all the different observables testing LU, including low-energy precision probes using pion, kaons and tau leptons~\cite{Bryman:2021teu}. In this work, we test LU between the second and third families of charged leptons through the ratio ($P=\pi,K$)~\cite{Marciano, DF,nosotros}
\eq{
R_{\tau / P}\equiv \frac{ \Gamma \left( \tau \to P \nu_{\tau}[\gamma] \right)}{\Gamma \left(P\to \mu \nu_{\mu}[\gamma]\right)}= \left\vert \frac{g_\tau}{g_\mu} \right\vert^2 R_{\tau/P}^{0}(1+\delta R_{\tau / P}), \label{MainDef}
}where $g_\mu=g_\tau$ according to LU, $\delta R_{\tau/P}$ denotes the radiative corrections, and $R_{\tau/P}^{(0)}$ is the leading order (tree-level) contribution given by
\eq{
R_{\tau/P}^{(0)}=\frac{1}{2}\frac{M_\tau^3}{m_\mu^2 m_P}\frac{(1-m_P^2/M_\tau^2)^2}{(1-m_\mu^2/m_P^2)^2}, \label{LO}
}which is free from hadronic couplings and quark mixing angles.

The first estimation for $\delta R_{\tau/P}$ including the complete set of real- and virtual-photon corrections was reported more than twenty-five years ago in ref.~\cite{DF}, where the values $\delta R_{\tau/\pi}=(0.16\pm0.14)\%$ and $\delta R_{\tau/K}=(0.90\pm0.22)\%$ were given. Nevertheless, there are important reasons to address this analysis again: the hadronic form factors modeled in ref.~\cite{DF} are different for real- and virtual-photon corrections, do not satisfy the correct QCD short-distance behavior, violate unitarity, analyticity, and the chiral limit at leading non-trivial orders; besides, a cutoff scheme was used to regulate the loop integrals, separating unphysically long- and short-distance corrections. Furthermore, the uncertainties given in ref.~\cite{DF} are unrealistic, being of the order of an expected purely $\mathcal{O}(e^2p^2)$ Chiral Perturbation Theory (ChPT) result (see e. g. the results in refs. \cite{CR1,CR2}).

Depending on the process at hand, different values of $\left|g_\tau/g_\mu\right|$ can be found in the literature:
\begin{enumerate}
\item $\Gamma(\tau \to P \nu_\tau  [\gamma]) / \Gamma(P \to\mu \nu_\mu[\gamma])$ ($P=\pi,K$). Using the values reported for $\delta R_{\tau/P}$ in ref.~\cite{DF}, the last HFLAV analysis~\cite{Amhis:2019ckw} quoted $\left|g_\tau/g_\mu\right|_\pi=0.9958\pm0.0026$ and $\left|g_\tau/g_\mu\right|_K=0.9879\pm0.0063$, at $1.6\sigma$ and $1.9\sigma$ of LU ($1.4\sigma$ and $2.0\sigma$ in ref.~\cite{Pich:2013lsa},  making use of the PDG input~\cite{Zyla:2020zbs}).
\item $\Gamma(\tau \to e \bar{\nu}_e \nu_\tau[\gamma])/\Gamma(\mu \to e \bar{\nu}_e \nu_\mu[\gamma])$. This purely leptonic extraction gives $\left|g_\tau/g_\mu\right|=1.0010\pm0.0014$~\cite{Amhis:2019ckw} ($1.0011\pm 0.0015$ in ref.~\cite{Pich:2013lsa}),  at $0.7\sigma$ of LU.
\item $\Gamma(W \to\tau \nu_\tau)/\Gamma(W \to\mu \nu_\mu)$. The weighted average of the recent $W$-boson decay determinations yields $\left|g_\tau/g_\mu\right|=0.995\pm 0.006$~\cite{Aad:2020ayz, CMS:2021qxj}, at $0.8\sigma$ of LU.
\end{enumerate}
Thus, a new estimation of $\delta R_{\tau/P}$ is timely to solve these discrepancies.

In this work, we present a new one-loop analysis of $\delta R_{\tau/P}$ by following a large-$N_C$ expansion where ChPT is enlarged by including the lightest resonances and respecting the high-energy behavior dictated by QCD~\cite{RChT}. On the one hand, $P$ decays are analyzed unambiguously by using the Standard Model (ChPT), being the estimation of the local counterterms the only model dependence: they are determined by matching ChPT with this large-$N_C$ expansion. On the other hand, $\tau$ decays must be scrutinized by using an effective approach encoding the hadronization of the QCD currents, so the model-independent contributions given by the point-like approximation are accompanied by the model- or  structure-dependent terms, which are calculated within the large-$N_C$ expansion quoted previously. 

Our new determination on $\delta R_{\tau/P}$ is used not only to test the LU in $\left|g_\tau/g_\mu\right|$, but also to revisit the CKM unitarity test via $\left|V_{us}/V_{ud}\right|$ in $\Gamma(\tau \to K \nu_\tau[\gamma])/\Gamma(\tau \to \pi \nu_\tau[\gamma])$ or through $\left|V_{us}\right|$ in $\Gamma(\tau \to K \nu_\tau[\gamma])$~\cite{Pich:2013lsa} and to update the constraints on possible non-standard interactions affecting this ratio~\cite{Cirigliano:2009wk, EFTtaudecays1,EFTtaudecays2}. Moreover, and as a by-product, we report the theoretical radiative corrections in individual $\tau \to P \nu_{\tau}[\gamma]$ decays.

This work is organized as follows. In sections \ref{tautoPnu} and \ref{Pmunu} we briefly review the real-photon and virtual-photon corrections to both $\tau\to P\nu_{\tau}[\gamma]$ and $P\to \mu \nu_{\mu}[\gamma]$
decays, respectively. Sections \ref{Resandapli} and \ref{appli} present, in turn, the results and the phenomenology associated to the observable $\delta R_{\tau/P}$. Finally, in section \ref{Conclu} we summarize the main conclusions of this work. For completeness, some relevant expressions and details of our computation are collected in appendix~\ref{avsd}. Appendix~\ref{DF-ours} shows a comparison between the results presented here and ref.~\cite{DF}. 


\section{The $\tau\to P\nu_{\tau}[\gamma]$ decay}\label{tautoPnu}

At the Born amplitude level the $\tau\to P\nu_{\tau}$ decay ($P= \pi,\, K$) is given by 
\eq{
\mathcal{M}_0=-G_F F_P V_{uD} M_\tau  \bar{u}(q)(1+\gamma_5)u(p_\tau),\label{Born-tPnu}
}where $G_F$ is the Fermi constant at leading-order, $f_P=\sqrt{2}F_P$ ($F_\pi\approx 92.2$ MeV and $F_K\approx 110$ MeV) is the meson decay constant, and $V_{uD}=V_{ud}\, (V_{us})$ is the CKM mixing element for the pion (kaon) case. 

At one-loop level, and in order to cancel the infrared divergences emerging from the calculation, it is required to consider the decay including the emission of a real photon, $\tau\to P\nu_{\tau}\gamma$. We follow here a fully photon-inclusive radiative decay,
\eq{
\tau\to P\nu_{\tau}[\gamma]\,\equiv\,\tau\to P\nu_{\tau}\,+\,\tau\to P\nu_{\tau}\gamma.}
Note that Lorentz invariance implies that higher order contributions are proportional to the lowest order amplitude given by eq. (\ref{Born-tPnu}). 

\subsection{Real-photon corrections}

\begin{figure*}[h!]
\begin{center}
\begin{tabular}{cccc}
\includegraphics[scale=0.58]{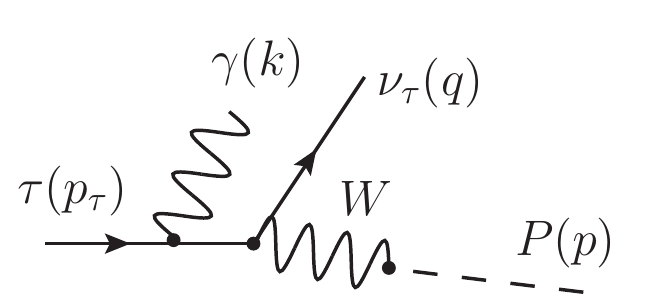} &
\includegraphics[scale=0.58]{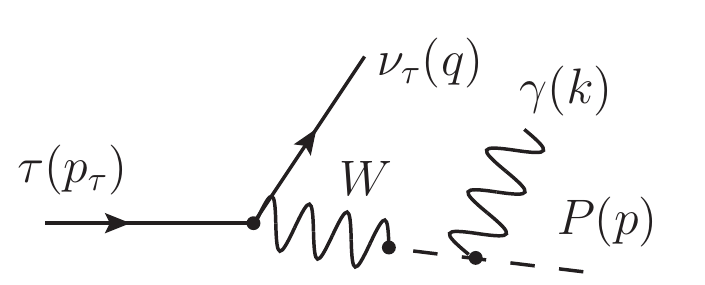} &
\includegraphics[scale=0.58]{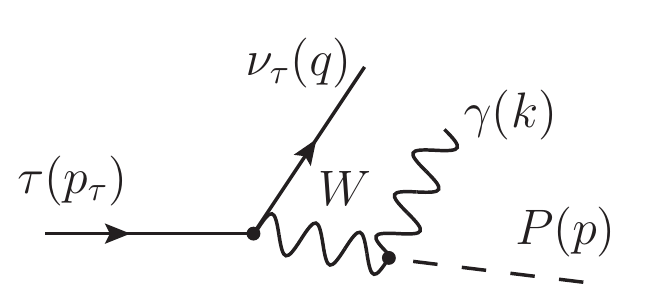} &
\\
{\small($a$)}& {\small($b$)}& {\small($c$)}& \\
&\includegraphics[scale=0.58]{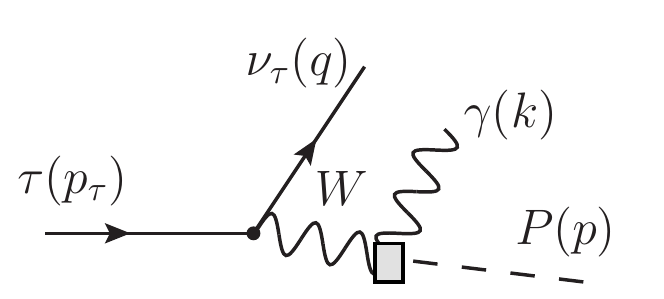}& &\\
&{\small($d$)}& &
\end{tabular}
\end{center}
\caption{Feynman diagrams for real-photon corrections in one-meson tau decays. The first three diagrams (a), (b), (c) stand for the inner bremsstrahlung contribution, which is model independent (structure-independent). The gray box in diagram (d) represents the structure-dependent contributions including the vector and axial currents. 
\label{radtauPnu}}
\end{figure*}

Let us start by considering the description of the real-photon corrections, for which, we follow refs.~\cite{Guo:2010dv, Guevara:2013wwa}. The matrix element can be split as
\begin{equation}
    \mathcal{M}[\tau^-(p_\tau)\to P^-(p)\nu_\tau(q)\gamma(k)]=\mathcal{M}_{IB_\tau}+\mathcal{M}_{IB_P}+\mathcal{M}_V+\mathcal{M}_A\,,
\end{equation}
where $ \mathcal{M}_{IB}=\mathcal{M}_{IB_\tau}+\mathcal{M}_{IB_P}$ comprises the model-independent (structure-independent) inner bremsstrahlung (IB) given by the radiation off the $\tau^-$ and off the $P^-$ meson. This part can be derived from the QCD low-energy theorems assuming an elementary point-like meson field with electromagnetic interaction dictated by the scalar QED Lagrangian~\cite{DF}. In this case,  the contribution of the seagull diagram (diagram (c) in figure \ref{radtauPnu}) is fixed by gauge invariance, in such a way that the sum of the diagrams (a), (b), and (c) in figure \ref{radtauPnu} is given by
\begin{equation}
 \mathcal{M}_{IB}=-iG_FV_{uD}eF_P M_\tau \Gamma_\mu \bar{u}(q)(1+\gamma_5)\left[\frac{2p^\mu}{2p\cdot k+k^2}+\frac{2p_\tau^\mu-k\!\!\!/\gamma^\mu}{-2p_\tau\cdot k+k^2}\right]u(p_\tau)\,,
\end{equation}
where $\Gamma_\mu=-\epsilon_\mu^*(k)$ for an on-shell photon~\cite{Guo:2010dv}. Alternatively, completely equivalent results are obtained using Chiral Perturbation Theory~\cite{ChPT}, the quantum effective field theory of QCD at low energies. Now, since the formalism preserves gauge invariance by construction, the seagull contribution is computed from the chiral Lagrangian.

On the other hand, diagram (d) includes the vector and axial structure-dependent contributions given by $\mathcal{M}_{V}+\mathcal{M}_{A}$, which yield~\cite{Guo:2010dv,Guevara:2013wwa}~\footnote{As already pointed out in ref.~\cite{Guo:2010dv}, there is a different convention for the definitions of the form factors involved in eqs. (\ref{eq:V}) and (\ref{eq:A}) with those reported in Decker and Finkemeier's (DF) work~\cite{Decker:1993ut}. Specifically, $F_V^P(W^2,k^2)^{DF}=\sqrt{2}m_P F_V^P(W^2,k^2)$, and $F_A^P(W^2,k^2)^{DF}=2\sqrt{2}m_P F_A^P(W^2,k^2)$. Both references also differ by a global $-i$ factor  which leads to a discrepancy for their estimations of the real-photon structure-dependent corrections ($\delta_{\textrm{rSD}}$) to $\tau\to P\nu_\tau$.   In this work,  we consider the expressions in ref.~\cite{Guo:2010dv} as the right ones since they were derived directly from the effective resonance lagrangian.\label{footnote1}}
\eq{
\mathcal{M}_V=&-G_FV_{uD}e\Gamma^\nu F_V^P(W^2,k^2)\epsilon_{\mu\nu\rho\sigma}k^\rho p^\sigma \bar{u}(q)\gamma^\mu(1-\gamma_5)u(p_\tau)\,\label{eq:V},\\
\mathcal{M}_A=&iG_FV_{uD}e\Gamma^\nu\Big\{F_A^P(W^2,k^2)\left[(W^2+k^2-m_P^2)g_{\mu\nu}-2k_\mu p_\nu\right] \nonumber\\
&-A_2^P(k^2)k^2g_{\mu\nu}+A_4^P(k^2)k^2(p+k)_\mu p_\nu\Big\}\bar{u}(q)\gamma^\mu(1-\gamma_5)u(p_\tau)\,, \label{eq:A}
}
where $W^2=(p_\tau-q)^2=(p+k)^2$ is the momentum carried by the $W$ propagator. At leading order in the chiral expansion, the form factors $A_2^P(k^2)$ and $A_4^P(k^2)$ are not independent and can be written in terms of a single form factor $B(k^2)$ (identical for $P=\pi,K$ at this order) as follows~\cite{Bijnens:1992en}:
\eq{
A_2(k^2)=-2B(k^2),\quad A_4(k^2)=-\frac{2B(k^2)}{W^2-m_P^2}.
}

\subsection{Vector and axial structure-dependent form factors}\label{sdff}
All the relevant model structure-dependent effects due to the QCD hadronization currents are encoded in the $F_V^P(W^2,k^2)$, $F_A^P(W^2,k^2)$, and $B(k^2)$ form factors. In this work, we take them from the Resonance Chiral Theory~\cite{RChT} computation in refs.~\cite{Guo:2010dv, Guevara:2013wwa}, where the resonance couplings contributing --upon integrating vector and axial mesons out-- up to the next-to-leading order chiral couplings in each parity sector were considered.  Regarding the vector form factors, they are given by~\footnote{Ideal $\omega-\phi$ mixing is assumed (expressions for the general case are given in refs.~\cite{Guo:2010dv, Guevara:2013wwa}).}
 
\eq{
F_V^{\pi}(W^2,k^2)&=\frac{1}{3F_\pi}\Bigg\{-\frac{N_C}{8\pi^2}+\frac{4F_V^2}{M_\rho^2-W^2}\frac{d_3(W^2+k^2)+d_{123}m_{\pi}^2}{M_\omega^2-k^2}\nonumber\\
&+\frac{2\sqrt{2}F_V}{M_V}\Bigg[\frac{c_{1256}W^2-c_{1235}m_\pi^2-c_{125}k^2}{M_\rho^2-W^2}+\frac{c_{1256}k^2-c_{1235}m_\pi^2-c_{125}W^2}{M_\omega^2-k^2}\Bigg]\Bigg\},\nonumber\\
\\
F_V^{K}(W^2,k^2)&=\frac{1}{F_K}\Bigg\{-\frac{N_C}{24\pi^2}+\frac{2F_V^2[d_3(W^2+k^2)+d_{123}m_K^2]}{M_{K^\star}^2-W^2}\Bigg[\frac{1}{M_\rho^2-k^2}+\frac{1}{3(M_\omega^2-k^2)}\nonumber \\
&-\frac{2}{3(M_\phi^2-k^2)}\Bigg]+\frac{2\sqrt{2}F_V}{3M_V}\frac{c_{1256}W^2-c_{1235}m_K^2-c_{125}k^2+24c_4\Delta_{K\pi}^2}{M_{K^\star}-W^2}\nonumber\\
&+\frac{\sqrt{2}F_V(c_{1256}k^2-c_{1235}m_K^2-c_{125}W^2)}{M_V}\Bigg[ \frac{1}{M_\rho^2-k^2}+{\frac{1}{3(M_\omega^2-k^2)}}-\frac{2}{3(M_\phi^2-k^2)}\Bigg]\Bigg\},
}
where $\Delta_{K\pi}^2\equiv m_K^2-m_\pi^2$, $F_V$ parametrizes the coupling of vector resonances to the vector current~\cite{RChT}, and $M_V$ is the mass of the lightest multiplet of resonances in the $3-$flavor symmetry limit [$M_V=M_\rho\, (M_{K^*})\sim 770$ (890) MeV for pion (kaon) case \cite{Guo:2009hi}]. Furthermore, the following combinations of coupling constants have been used~\cite{Dumm:2012vb}~\footnote{Their values will be discussed later, in light of structure-dependent constraints.}
\eq{
c_{125}&=c_1-c_2+c_5,\nonumber\\
c_{1256}&=c_1-c_2-c_5+2c_6,\nonumber\\
c_{1235}&=c_1+c_2+8c_3-c_5,\nonumber\\
d_{123}&=d_1+8d_2-d_3.
}
in terms of the linear combinations of $c_i$ and $d_j$ couplings introduced (further details are found in ref.~\cite{RuizFemenia:2003hm}).

As far as the axial form factors are concerned, we start the discussion by recalling that
\eq{\label{Bformfactor}
B(k^2)=F_P\frac{F_V^{\pi\pi}|_{I=1}-1}{k^2}  =\frac{F_P}{M_V^2-k^2}
\,,    
}
with $F_V^{\pi\pi}|_{I=1}=F_V^{KK}|_{I=1}$ neglecting isospin breaking effects~\cite{Gonzalez-Solis:2019iod} and $M_V=M_\rho$. The remaining axial form factors are
\eq{
F_A^{\pi}(W^2,k^2)=& \frac{F_V}{2F_{\pi}}\frac{F_V-2G_V}{M_\rho^2-k^2}-\frac{F_A}{2F_\pi}\frac{F_A}{M_{a_1}^2-W^2}+\frac{\sqrt{2}}{F_\pi}\frac{F_AF_V}{M_{a_1}^2-W^2}\frac{\lambda_0 m_\pi^2-\lambda'k^2-\lambda''W^2}{M_\rho^2-k^2},
}
and
\eq{\label{eq:FAK}
F_A^{K}(W^2,k^2)=&-\frac{F_A}{2F_K}\frac{F_A}{M_{K_1}^2-W^2}
+\left[\frac{\sqrt{2}F_A F_V}{2F_K}\frac{\lambda_0 m_K^2-\lambda'k^2-\lambda''W^2}{M_{K_1}^2-W^2}+\frac{F_V
\left(F_V-2G_V\right)}{4F_K}\right]
\nonumber\\
&\times \left(\frac{1}{M_\rho^2-k^2}+\frac{1}{3}\frac{1}{M_\omega^2-k^2}+\frac{2}{3}\frac{1}{M_\phi^2-k^2}\right).
}
In addition to $F_A$ $(G_V)$, giving the coupling of an axial (vector) resonance to the axial (vector) current~\cite{RChT}, we used~\cite{GomezDumm:2003ku}
\eq{
\sqrt{2}\lambda_0 =-\left(4\lambda_1+\lambda_2+\frac{\lambda_4}{2}+\lambda_5\right),\quad
\sqrt{2}\lambda' =\left(\lambda_2-\lambda_3+\frac{\lambda_4}{2}+\lambda_5\right),\quad
\lambda''=\lambda'+\frac{\lambda_3}{\sqrt{2}}.
}
In the form factors above, resonance propagators are short for the corresponding expressions including (energy-dependent) off-shell widths, {\it i.e.} $(M_R^2-x)^{-1}\to(M_R^2-x-iM_R\Gamma_R(x))^{-1}$. For the case of the $\rho(770)$ meson this width is given by ref.~\cite{GomezDumm:2000fz} ($\sigma_P(x)=\sqrt{1-4m_P^2/x}$)
\begin{equation}
\Gamma_\rho(x)=\frac{xM_\rho}{96\pi F_\pi^2}\left[\theta(x-4m_\pi^2)\sigma_\pi^3(x)+\frac{1}{2}\theta(x-4m_K^2)\sigma_K^3(x)\right]\,,
\end{equation}
with analogous expression for the $K^*(892)$~\cite{Jamin:2006tk}~\footnote{Here $\eta_8$ was identified with the physical $\eta$ meson for simplicity. The induced error is negligible.}
\begin{equation}
\Gamma_{K^*}(x)=\frac{xM_{K^*}}{256\pi F_\pi^2}\Big[\theta\left(x-(m_\pi+m_K)^2\right)\sigma_{K\pi}^3(x)+\theta\left(x-(m_K+m_\eta)^2\right)\sigma_{K\eta}^3(x)\Big]\,,
\end{equation}
where $\sigma_{PQ}(x)=2q_{PQ}(x)/\sqrt{x}$ and 
\begin{equation}
q_{PQ}(x)=\frac{\sqrt{\left(x-(m_P+m_Q)^2\right)\left(x-(m_P-m_Q)^2\right)}}{2\sqrt{x}} \,.
\end{equation}
A constant width will be used for the very narrow $\omega(782)$ and $\phi(1020)$ mesons, given by the PDG value~\cite{Zyla:2020zbs}.

Off-shell $a_1(1260)$ width was computed in refs.~\cite{Dumm:2009va, Nugent:2013hxa}. It includes the $3\pi$~\cite{Dumm:2009va} and $KK\pi$~\cite{Dumm:2009kj} cuts.
Finally, the $K_1$ resonance will be accounted for following ref.~\cite{Guo:2008sh}
\begin{equation}
\Gamma_{K_{1\,L/H}}(x)= \Gamma_{K_{1\,L/H}}\frac{x}{M_{K_{1\,L/H}}^2}\left[\frac{\sigma_{K\rho}^3(x)+\sigma_{K^*\pi}^3(x)}{\sigma_{K\rho}^3(M_{K_{1\,L/H}}^2)+\sigma_{K^*\pi}^3(M_{K_{1\,L/H}}^2)}\right]\,,
\end{equation}
with the mixing scheme explained in this reference. To account for it, in eq. (\ref{eq:FAK}), $(M_{K_1}^2-W^2)^{-1}\to\mathrm{cos}^2\theta_A(M_{K_{1H}}^2-W^2)^{-1}+\mathrm{sin}^2\theta_A(M_{K_{1L}}^2-W^2)^{-1}$, with mixing angle $\theta_A=[37,58]^\circ$ and the light and heavy states with masses $1270$ and $1400$ MeV, respectively.

The couplings appearing in the $F_{V/A}^{\pi,K}(W^2,k^2)$~\footnote{The $B(k^2)$ form factor has an appropriate  high-energy behaviour, as the dispersive $F_V^{\pi\pi}|_{I=1}$ form factor~\cite{Dumm:2013zh,Gonzalez-Solis:2019iod} complies with the Brodsky-Lepage limit~\cite{Brodsky-Lepage}.} form factors are not restricted by chiral symmetry. Fortunately, some of them are predicted by demanding the behaviour given by the operator product expansion of QCD to the relevant Green functions~\cite{ RChT, RuizFemenia:2003hm, Weinberg:1967kj,Cirigliano:2004ue, Kampf:2011ty, Roig:2013baa}.

We will consider two different approaches for this:

\textbf{\textit{Scenario (a)}}: We consider the two-pion vector form factor and the (axial-)vector two-point Green functions. Then, the short-distance relations determine
\begin{equation} \label{constr_sca1}
F_V G_V = F_{\pi}^2\,,\quad F_V^2-F_A^2=F_\pi^2\,,\,\quad F_V^2 M_V^2=F_A^2M_A^2 \,.
\end{equation}
As a first approach to our form factors, we will neglect the contribution from all other operators (which is suppressed in the chiral expansion) and set all $\lambda_i$, $c_j$ and $d_k$ couplings to zero. Within this approach, demanding the QCD ruled short-distance behaviour to the $\pi$ axial form factor (giving the $\pi\to\gamma$ matrix element) determines~\cite{RChT}
\begin{equation}\label{eq:1stapprox}
F_V=\sqrt{2}F_\pi\,,\quad G_V=\frac{F_\pi}{\sqrt{2}}\,,\quad F_A=F_\pi\,.  
\end{equation}
The relation $M_A=\sqrt{2}M_V$, coming from the 2nd Weinberg sum rule \cite{Weinberg:1967kj} (the last constraint in (\ref{constr_sca1})) and these relations, should be understood as applying to the lowest-lying large-$N_C$ axial mass when all other multiplets (an infinite number of them for $N_C\to\infty$ \cite{tHooft:1973alw}) are neglected, and does not correspond to the physical value of the $a_1/K_1$ mass \cite{Masjuan:2007ay}, that we will use instead. Using these relations (and all other couplings vanishing), we will employ the form factors 
\begin{equation}\label{scenarioA}
\bar{F}_V^{P}(W^2,k^2)=\bar{F}_V^{P}=\frac{-N_C}{24\pi^2F_P}\,,\quad  \bar{F}_A^{P}(W^2,k^2)=\bar{F}_A^{P}(W^2)=\frac{-F_A^2}{2F_P(M_A^2-W^2)}\,,
\end{equation}
as first approximation (which we indicate with a bar in these form factors), 
where $M_V=M_\rho(M_{K^*})$ and $M_A=M_{a_1}(M_{K_1})$ for $P=\pi(K)$.

\textbf{\textit{Scenario (b)}}: In a more refined analysis (in which the form factors will be tilded) we take into account the short-distance constraints corresponding to the analysis of the two and three-point Green functions~\cite{RChT, RuizFemenia:2003hm, Cirigliano:2004ue, Kampf:2011ty, Roig:2013baa}, yielding, for the couplings between a vector and an axial resonance
\begin{equation}\label{eq:lambdas}
\lambda_0=\frac{\lambda'+\lambda''}{4}\,,\quad \lambda'=\frac{F_V}{2\sqrt{2}\sqrt{F_V^2-F_{\pi}
^2}}\,\quad \lambda''=\frac{2F^2-F_V^2}{2\sqrt{2}F_V\sqrt{F_V^2-F_{\pi}^2}}\,.
\end{equation}
Note that these expressions simplify when the constraint $F_V=\sqrt{3}F_{\pi}
$~\cite{Roig:2013baa} (coming from the odd-intrinsic parity sector) is used. In fact, it modifies the relation (\ref{eq:1stapprox}) to
\begin{equation}\label{eq:refinedapprox}
F_V=\sqrt{3}F_\pi\,,\quad G_V=\frac{F_\pi}{\sqrt{3}}\,,\quad F_A=\sqrt{2}F_\pi\,.  
\end{equation}
Again the relation $M_A=\sqrt{3/2}M_V$, coming from the 2nd Weinberg sum rule~\cite{Weinberg:1967kj} and these relations, does not apply here, since we consider the physical values. 

The consistent set of short-distance constraints in the odd-intrinsic parity sector includes~\cite{Kampf:2011ty, Roig:2013baa}
\begin{equation}\label{eq:ConsistentSetOdd}
    c_{125}=0,\;c_{1235}=0,\;c_{1256}=-\frac{N_CM_V}{32\sqrt{2}\pi^2F_V},\;d_{123}=\frac{F^2}{8F_V^2},\;d_3=-\frac{N_CM_V^2}{64\pi^2F_V^2},\;F_V=\sqrt{3}F_{\pi} \,.
\end{equation}
Using the above equations, the $F_{V}^P(W^2,k^2)$ form factors read
\eq{
\tilde{F}_V^{\pi}(W^2,k^2)&=\frac{1}{3F_\pi}\Bigg\{-\frac{N_C}{8\pi^2}+\frac{4F_V^2}{M_\rho^2-W^2}\frac{d_3(W^2+k^2)+d_{123}m_\pi^2}{M_\omega^2-k^2}\nonumber\\
&+\frac{2\sqrt{2}F_V}{M_V}\Bigg[\frac{c_{1256}W^2}{M_\rho^2-W^2}+\frac{c_{1256}k^2}{M_\omega^2-k^2} \Bigg]\Bigg\}
}
and
\eq{
\tilde{F}_V^{K}(W^2,k^2)=&\frac{1}{F_K}\Bigg\{-\frac{N_C}{24\pi^2}+\frac{2F_V^2\left[d_3(W^2+k^2)+d_{123}   m_K^2\right]}{M_{K^\star}^2-W^2}\Bigg(\frac{1}{M_\rho^2-k^2}+\frac{1}{3(M_\omega^2-k^2)}\nonumber\\
&-\frac{2}{3(M_\phi^2-k^2)}\Bigg)+\frac{2\sqrt{2}F_V}{3M_V}\frac{c_{1256}W^2+24c_4\Delta_{K\pi}^2}{M_{K^\star}^2-W^2}\nonumber\\
&+\frac{\sqrt{2}F_V(c_{1256}k^2)}{M_V}\left(\frac{1}{M_\rho^2-k^2}+\frac{1}{3(M_\omega^2-k^2)}-\frac{2}{3(  M_\phi^2-k^2)}\right)\Bigg\},
}
with $F_V$, $c_{1256}$, $d_{123}$ and $d_3$ given by eqs. (\ref{eq:ConsistentSetOdd}). In this way, all relevant couplings are restricted by short-distance constraints but $c_4$. For this we will take $c_4=-0.0024\pm0.0006$~\cite{Chen:2013nna}
(see related discussions in ref.~\cite{Guevara:2016trs} and references therein).
Finally, and considering (\ref{eq:ConsistentSetOdd}), the chiral and $U(3)$ flavor limits of $\tilde{F}^{\pi,K}$ reads

\begin{equation}\label{scenarioB_vectorpart}
\tilde{F}^{\pi,K}_V(W^2,k^2)=\frac{-N_C M_V^4}{24\pi^2F_P(k^2-M_V^2)(W^2-M_V^2)}\,.
\end{equation}
The common correction to the chiral limit result amounts to add to the numerator of the previous expression $+4\pi^2F_P^2m_P^2$. There is an additional correction for the kaon case, with a contribution of $\frac{16\sqrt{6}c_4\Delta^2_{K\pi}}{M_V(M_V^2-W^2)}$ to $\tilde{F}^K_V(W^2,k^2)$.
\\

The $\tilde{F}_{A}^P(W^2,k^2)$ form factors also simplify using eqs.~(\ref{eq:lambdas}) and (\ref{eq:refinedapprox}). Their chiral and $U(3)$ symmetry limits (that will be used in the virtual-photon corrections) read
\begin{equation} \label{scenarioB_axialpart}
\tilde{F}_{A}^P(W^2,k^2)=\frac{F_P}{2}\left[\frac{1}{M_V^2-k^2}-\frac{2}{M_A^2-W^2}+\frac{W^2-3k^2}{(M_V^2-k^2)(M_A^2-W^2)}\right]\,.
\end{equation}

\subsection{Virtual-photon corrections}

Now, analogously to the $\tau\to P\nu_\tau \gamma$ decays, it turns out natural to study the virtual-photon corrections to $\tau\to P\nu_\tau$ decays by separating their contributions into the  structure-dependent and structure-independent parts.  

\subsubsection{Virtual-photon structure-independent corrections}

\begin{figure*}[h!]
\begin{center}
\begin{tabular}{ccc}
\includegraphics[scale=0.6]{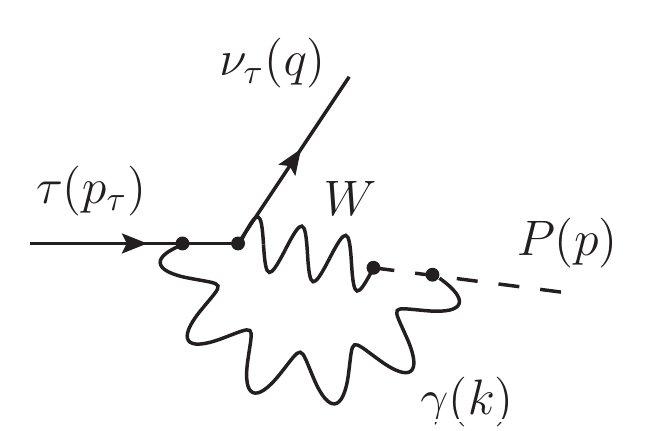} &
\includegraphics[scale=0.6]{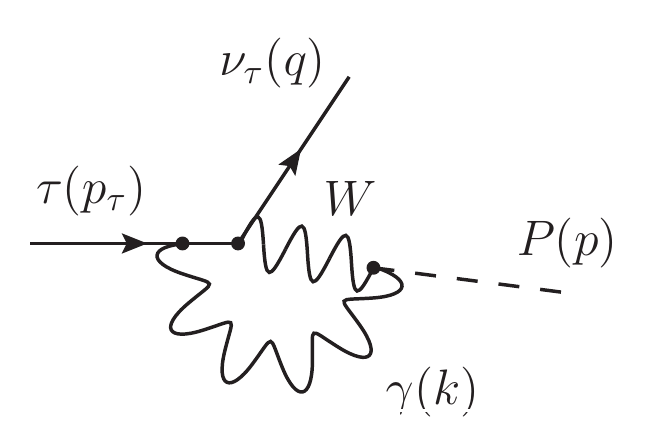} &
\includegraphics[scale=0.6]{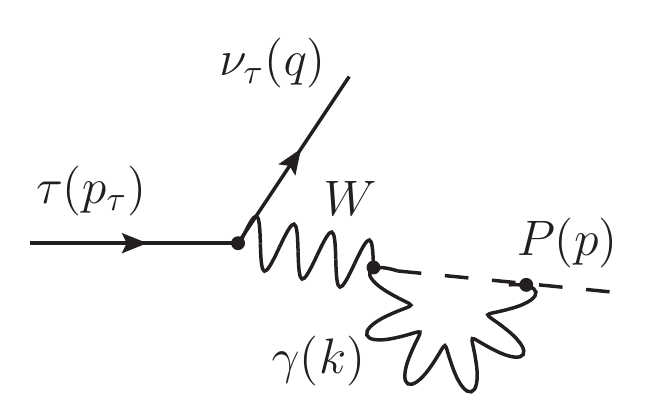} \\
($a$)& ($b$)& ($c$) \\
\includegraphics[scale=0.6]{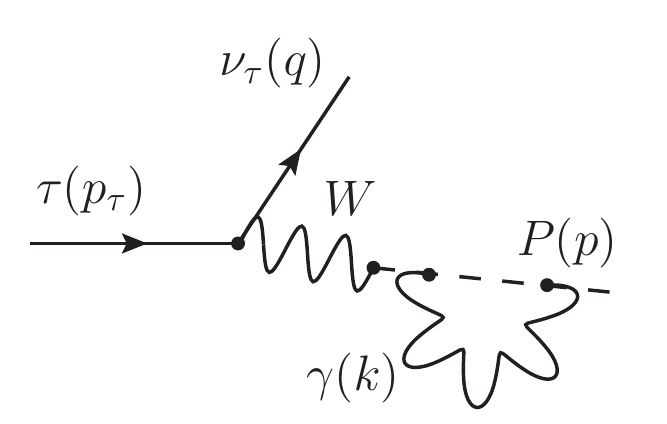} &
\includegraphics[scale=0.6]{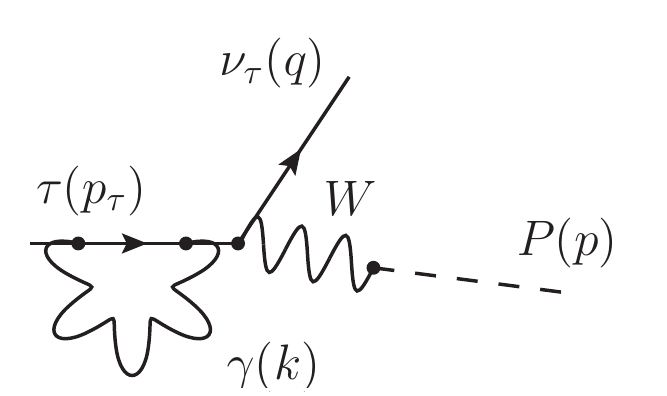} &
\includegraphics[scale=0.6]{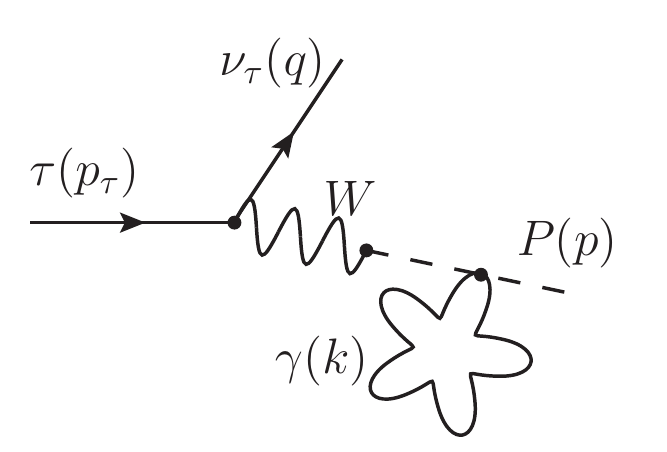}\\
($d$)& ($e$)& ($f$)\\ 
\end{tabular}
\end{center}
\caption{Structure-independent or point-like meson corrections to $\tau^-\to P^-\nu_\tau$. Notice that the contribution of diagram $(f)$ vanishes after the meson mass renormalization. 
A similar set of diagrams also describes the virtual corrections to the $P\to \mu\nu_\mu$ decays with the proper replacements dictated by  crossing symmetry.}\label{sivc}
\end{figure*}

The virtual-photon structure-independent (vSI), also called point meson loops correction (PML) in ref.~\cite{DF}, to $\tau\to P\nu_\tau$ are given by the Feynman diagrams depicted in figure \ref{sivc}. By defining the ratio of each diagram in figure \ref{sivc} with the Born amplitude in eq. (\ref{Born-tPnu}) as $\delta\mathcal{M}_{i}\equiv \mathcal{M}_{i}/\mathcal{M}_0$ where $i=(a), (b), ..., (e)$, we have that  
\eq{\label{vIB}
\delta\mathcal{M}_{a}(m_\ell^2,m_P^2,m^2)&=\frac{\alpha}{4\pi}\left[3B_0^P+B_1^P-2m_P^2 C_1+2m_{\ell}^2C_0-2B_0^{\ell}-\frac{20}{5}B_1^{\ell}\right],\nonumber\\
\delta\mathcal{M}_{b}(m_\ell^2,m_P^2,m^2)&=\frac{\alpha}{4\pi}\left[1+2(B_1^\ell-B_0^\ell)\right],\nonumber\\
\delta\mathcal{M}_{c}(m_\ell^2,m_P^2,m^2)&=\frac{\alpha}{4\pi}\left[-2B_0^P-B_1^P\right],\nonumber\\
\delta\mathcal{M}_{d}(m_\ell^2,m_P^2,m^2)&=\frac{\alpha}{4\pi}\left[\frac{B_0^P}{2}-B_1^P+m_P^2(B_0'^P-B_1'^P) \right],\nonumber\\
\delta\mathcal{M}_{e}(m_\ell^2,m_P^2,m^2)&=\frac{\alpha}{4\pi}\left[\frac{1}{2}+B_1^\ell+m_\ell^2(4B_0'^\ell+2B_1'^\ell)\right]\,,
}and the total contribution is given by $\delta\mathcal{M}_{\mathrm{vSI}}\equiv \sum\limits_{i}\delta\mathcal{M}_{i}
$:

\eq{\label{OurResult-RadCor-tau_p2PointLikePION}
\delta \mathcal{M}_{\mathrm{vSI}}=\frac{\alpha}{4\pi}\Bigg\{\frac{3}{2}\!+\!\frac{3}{2}B_0^P\!-\!B_1^P\!+\!m_P^2(B_0^{\prime P}\!-\!B_1^{\prime P}\!-\!2C_1)\!-\!4B_0^{\ell}\!-\!B_1^{\ell} \!+\!m_{\ell}^2(B_0^{\prime \ell}\!+\!2B_1^{\prime\ell}\!+\!2C_0)\Bigg\},
}
where $B_i^{\ell,P}$, $B_i'^{\ell,P}$ and $C_i$ are the corresponding two- and three-point Passarino-Veltman loop functions,
\begin{eqnarray}
B_i^{\ell,P}&=&B_i(m_{\ell,P}^2,m_{\ell,P}^2,m^2) \,, \nonumber \\
B_i'^{\ell,P}&=&\left.\frac{\partial B_i(p^2,m_{\ell,P}^2,m^2)}{\partial p^2}\right|_{p^2=-m_{\ell,P}^2} \,, \nonumber \\
C_i&=&C_i(m_P^2,m_\ell^2,0,m_P^2,m^2,m_\ell^2) \,.
\end{eqnarray}

 In the above expressions, $m$ is associated to a photon mass that regulates the IR divergences due to the photon propagator. Nevertheless, such divergent behaviour cancels by adding to these virtual corrections the integrated rate for internal bremsstrahlung~\cite{DF}.

The above expression can be taken to the analytical form ($r_\ell\equiv m_\ell/m_P$ and $\mu$ is the dimensional regularization mass scale)
\begin{eqnarray}\label{deltaMPV}
\delta\mathcal{M}_{\mathrm{vSI}}(m_\ell^2,m_P^2,m^2)=\frac{\alpha}{4\pi}\Bigg\{-\frac{3}{2}\Delta+\frac{3}{2}\mathrm{log}\frac{m_P^2}{\mu^2}-4-4\left[\frac{1+r_\ell^2}{1-r_\ell^2}\mathrm{log}\,r_\ell+1\right]\mathrm{log}\frac{m}{m_P}\nonumber\\
+2\frac{1+r_\ell^2}{1-r_\ell^2}(\mathrm{log}\,r_\ell)^2+\left(5-\frac{4r_\ell^2}{1-r_\ell^2}\right)\mathrm{log}\,r_\ell\Bigg\}\,,
\end{eqnarray}
where $\Delta=2\mu^{D-4}/(4-D) -\gamma_{\mathrm{Euler}}+\log 4\pi$ captures the divergence in dimensional regularization. It is important to mention that even though some of the individual contributions in eq. (\ref{vIB}) are expressed in a different way than those ones in ref.~\cite{DF}, we completely agree with the total contribution given in eq. (\ref{deltaMPV}).

\subsubsection{Virtual-photon structure-dependent corrections} \label{vSDtau}
\begin{figure}
\begin{center}
\includegraphics[scale=0.7]{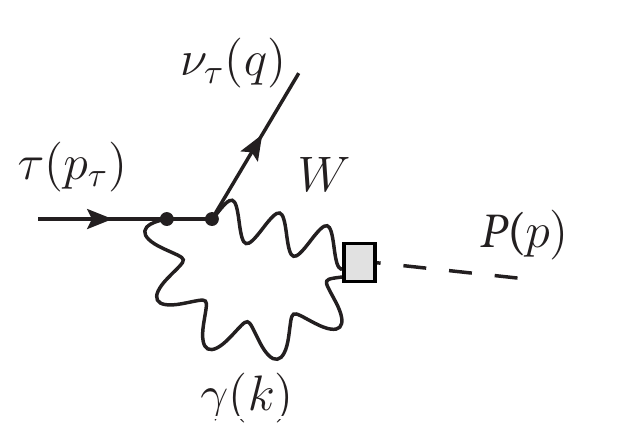} 
\end{center}
\caption{Feynman diagram corresponding to the virtual-photon structure-dependent (vSD) contributions to $\tau \to P \nu_\tau$ decays. The gray shaded box stands for the structure-dependent vector and axial hadronic currents. 
\label{feynman}}
\end{figure}
Now, let us focus on the virtual-photon structure-dependent contributions (vSD). The decays with an off-shell photon presented in section \ref{sdff} define the form factors entering the loops of the vSD contributions and the relevant diagram where one photon vertex is attached to the $\tau$ lepton (see figure \ref{feynman}), reads 
\eq{
i\mathcal{M}[\tau \to P\nu_\tau]|_{\mathrm{vSD}}=&
G_F V_{uD}e^2 \!\int\! \frac{\textrm{d}^dk}{(2\pi)^d} \frac{\ell^{\mu\nu}}{k^2[(p_\tau\!+\!k)^2\!-\!M_\tau^2]}\times\nonumber\\
& \left[i \epsilon_{\mu\nu\lambda\rho}k^\lambda p^\rho F_V^P (W^2\!,k^2)
\!+\!F_A^P(W^2\!,k^2)\lambda_{1\mu\nu}\!+\!2 B(k^2) \lambda_{2\mu\nu}\! \right]\!,\label{amp-virsd} 
}
with the definitions
\eq{
\ell^{\mu\nu}&=\bar{u}(q)\gamma^\mu(1-\gamma_5)[(\cancel{p}_\tau+\cancel{k})+M_\tau]\gamma^\nu u(p_\tau),\nonumber \\
\lambda_{1\mu\nu}&=\left[(p+k)^2+k^2-m_P^2\right]g_{\mu\nu}-2k_{\mu}p_{\nu},\nonumber \\
\lambda_{2\mu\nu}&=k^2 g_{\mu\nu}-\frac{k^2 (p+k)_{\mu}p_{\nu}}{(p+k)^2-m_P^2}.
}
After the loop integration, the amplitude for the vSD contributions can be conveniently written as follows
\eq{
\mathcal{M}[\tau \to P\nu_\tau]|_{\mathrm{vSD}}=&\frac{\alpha}{4\pi}G_F V_{uD} \left(f_{V}+f_{A}+f_B\right)\, M_\tau\, \bar{u}(q)(1+\gamma_5)u(p_\tau) \label{amp-virsd-2},
}
where the $f_V$, $f_{A}$, and $f_{B}$ functions include the relevant one-loop contributions coming from the $F_V(W^2,k^2)$, $F_A(W^2,k^2)$, and $B(k^2)$ form factors appearing in the integral of eq. (\ref{amp-virsd}), respectively. Then, analogously to the previous section by defining $\delta\mathcal{M}_{\textrm{vSD}}\equiv \mathcal{M}_\textrm{vSD}/\mathcal{M}_0$, we have that:
\eq{
\delta\mathcal{M}_{\textrm{vSD}}=\frac{\alpha}{4\pi F_P}\left(f_V+f_A+f_B\right).
}
In appendix \ref{avsd} we present all the relevant expressions in the two scenarios studied in section \ref{sdff}. 

\subsection{Total contributions}

The total decay rate can be organized as:
\eq{
\Gamma_{\tau_{P2[\gamma]}}= 
\Gamma^{(0)}_{\tau_{P2}}
\, S_{\rm EW}
\Bigg\{ 1 + \frac{\alpha}{\pi}  \,  G (m_P^2/M_\tau^2)  \Bigg\}
\Bigg\{  1 - \frac{3\alpha}{2\pi}  \log \frac{m_\rho}{M_\tau} +  \delta_{\tau P}\big|_{\mathrm{rSD}} +  \delta_{\tau P}\big|_{\mathrm{vSD}}\Bigg\} \,,
\label{eq:indratetau}
}
where $\Gamma^{(0)}_{\tau_{P2}}$ is the rate in absence of radiative corrections, 
\begin{equation}
\Gamma^{(0)}_{\tau_{P2}}=\frac{G_F^2 |V_{uD}|^2F_P^2 }{8\pi} M_\tau^3\left(1-\frac{m_P^2}{M_\tau^2}\right)^2, 
\label{eq:indratetauLO}
\end{equation}
being $D=d,s$ for $P=\pi,K$, respectively. $S_{\rm EW}=1.0232\simeq  1 + \frac{2 \, \alpha}{\pi}  \log \frac{m_Z}{m_\rho} $ corresponds to the (universal) leading short-distance electroweak correction~\cite{Marciano} and it cancels in the ratio  $R_{\tau/P}$. The first bracketed term captures the universal long-distance pointlike correction and will be reported later. In turn, the structure-dependent contributions have been split into the real-photon (rSD) and virtual-photon (vSD) corrections coming from the amplitudes in eqs.(\ref{eq:V}), (\ref{eq:A}) and (\ref{amp-virsd-2}), respectively. 

\section{The $P\to \mu \nu_{\mu}[\gamma]$ decay}\label{Pmunu}

Regarding the real-photon corrections to $P_{\mu2}$ decays, they are completely analogous to the radiative tau decays discussed before. Defining $P^-(p)\to\mu^-(p')\gamma(k)\bar{\nu}_\mu(q)$, the matrix element for the model-independent inner bremsstrahlung ($IB$) as well as for the vector and axial structure-dependent parts are given by
\eq{
 \mathcal{M}_{IB}&=-iG_FV_{uD}eF_P m_\mu \Gamma_\mu \bar{u}(q)(1+\gamma_5)\left[\frac{2p^\mu}{2p\cdot k+k^2}+\frac{2p'^\mu-k\!\!\!/\gamma^\mu}{-2p'\cdot k+k^2}\right]v(p')\,,\\
\mathcal{M}_V&=G_FV_{uD}e\Gamma^\nu F_V(\overline{W}^2,k^2)\epsilon_{\mu\nu\rho\sigma}k^\rho p^\sigma \bar{u}(q)\gamma^\mu(1-\gamma_5)v(p')\,\label{MV-P2},\\
\mathcal{M}_A&=-iG_FV_{uD}e\Gamma^\nu\Big\{F_A(\overline{W}^2,k^2)\left[(\overline{W}^2+k^2-m_P^2)g_{\mu\nu}-2k_\mu p_\nu\right]\nonumber\\
&-A_2(k^2)k^2g_{\mu\nu}+A_4(k^2)k^2(p+k)_\mu p_\nu\Big\}\bar{u}(q)\gamma^\mu(1-\gamma_5)v(p')\,,\label{MA-P2}
}
where $\overline{W}^2=(p'+q)^2=(p-k)^2$. Now, crossing symmetry evidences the relation to the $\tau^-\to P^-\gamma \nu_\tau$ processes. Thus, the form factors in eqs. (\ref{MV-P2}) and (\ref{MA-P2})  are the same than those in (\ref{eq:V}) and (\ref{eq:A}) by replacing $W^2\to\overline{W}^2$ (the $A_{2,4}(k^2)$ functions are not affected). 

Contrary to $\tau_{\mu 2[\gamma]}$, inclusive $P_{\mu 2[\gamma]}$ decay rate can be analyzed unambiguously within the Standard Model (Chiral Perturbation Theory), being the estimation of the local counterterms the only model dependence. Therefore, and except for the estimation of the local counterterms, only the low-energy limit of the form factors given in the previous section are required. Similarly to (\ref{eq:indratetau}), we follow the notation proposed in ref.~\cite{Marciano} and the numbers reported in refs.~\cite{CR1,CR2}:\eq{
&\Gamma_{P_{\mu 2 [\gamma]}} = 
\Gamma^{(0)}_{P_{\mu 2 }} \!
\, S_{\rm EW} \Bigg\{ 1 + \frac{\alpha}{\pi}  \,  F( m_\mu^2/m_P^2)  \Bigg\}
\Bigg\{  1 - \frac{\alpha}{\pi}   \Bigg[ \frac{3}{2}  \log \frac{m_\rho}{m_P}\nonumber\\
&+   c_1^{(P)}  + \frac{m_\mu^2}{m_\rho^2}   \bigg(c_2^{(P)}  \, \log \frac{m_\rho^2}{m_\mu^2}   +  c_3^{(P)} 
+ c_4^{(P)} (m_\mu/m_P) \bigg)  -  \frac{m_P^2}{m_\rho^2} \,  \tilde{c}_{2}^{(P)}  \, \log \frac{m_\rho^2}{m_\mu^2}  \Bigg] \Bigg\} \,,
\label{eq:indrate}
}
where again $S_{\rm EW}=1.0232\simeq  1 + \frac{2 \, \alpha}{\pi}  \log \frac{m_Z}{m_\rho} $ corresponds to the (universal) leading short-distance electroweak correction~\cite{Marciano}. The first bracketed term is the universal long-distance correction (point-like approximation, originally calculated in ref.~\cite{Kinoshita:1959ha} and to be given later), the second bracketed term includes the structure-dependent contributions and $\Gamma^{(0)}_{P_{\mu 2 }}$ is the rate in absence of radiative corrections,
\begin{equation}
\Gamma^{(0)}_{P_{\mu 2}} \,=\,  \frac{G_F^2 |V_{uD}|^2  F_P^2 }{4 \pi} \, 
m_P  \, m_\mu^2  \, \left(1 - \frac{m_\mu^2}{m_P^2} \right)^2 \,, 
\end{equation}
being $D=d,s$ for $P=\pi,K$, respectively. The numerical values for $c_n^{(P)}$ are reported in table~\ref{tab:tab1}~\cite{CR1,CR2}. Note that the most important uncertainties come from the estimations of the local counterterms, which were computed  
by matching ChPT and the large-$N_C$ expansion including resonances quoted previously, that is, by using the resonance form factors explained in the previous section.  
\\
\begin{table}[t!]
\begin{center}
\begin{tabular}{|c|c|c|}
\hline
  & $(P=\pi)$  & $(P=K)$    \\[5pt]
\hline 
$c_{1}^{(P)}$ & $-2.56\pm0.5_{\rm m }$ & $-1.98\pm0.5_{\rm m }$ \\
 $c_2^{(P)}$  & $5.2 \pm 0.4_{L_9} \pm 0.01_\gamma$  &  $4.3 \pm 0.4_{L_9} \pm 0.01_\gamma $ \\
 $c_3^{(P)}$   
&   
$ -10.5 \pm 2.3_{\rm m } \pm 0.53_{L_9} 
$
& 
$ -4.73 \pm 2.3_{ \rm m} \pm 0.28_{L_9}
$ 
\\
 $c_4^{(P)} $  &
$1.69 \pm 0.07_{L_9} $
&  
$ 0.22 \pm 0.01_{L_9} $ 
\\
$\tilde{c}_2^{(P)}$  &   0  &  $ (7.84 \pm 0.07_\gamma) \cdot 10^{-2}  $ \\
\hline
\end{tabular}
\end{center}
\caption{Numerical values for $c_n^{(P)}$ of (\ref{eq:indrate})~\cite{CR1,CR2} ($c_{1}^{(P)}$ from ref.~\cite{DescotesGenon:2005pw}). The uncertainties correspond to the input values  $L_9^r (\mu=m_\rho) = (6.9 \pm 0.7) \cdot 10^{-3} $, $\gamma = 0.465 \pm 0.005$, and to the estimation of the counterterms (${\rm m}$, from matching), affecting only $c_1^{(P)}$ and $c_3^{(P)}$.}
\label{tab:tab1}
\end{table}


\section{Determination of $\delta R_{\tau/P}$}
\label{Resandapli}

As it has been spotlighted in the introduction, the main aim of this work is to determine $R_{\tau / P}$,
\eq{
R_{\tau / P}\equiv \frac{ \Gamma \left( \tau \to P \nu_{\tau}[\gamma] \right)}{\Gamma \left(P\to \mu \nu_{\mu}[\gamma]\right)}=  R_{\tau/P}^{0}(1+\delta R_{\tau / P})=R_{\tau/P}^{0}(1+ \delta_{\tau P} - \delta_{P \mu } ) , \label{MainDef2}
}
where $R_{\tau/P}^{(0)}$ is the leading order contribution given in (\ref{LO}) and $\delta R_{\tau/P}$ captures the radiative corrections. For some contributions it has been convenient to split the radiative corrections into the contributions from the $\tau$ and $P$ decays, $\delta_{\tau P}$ and $\delta_{P \mu }$ respectively.

\subsection*{Structure-independent contributions}

Considering the structure-independent (SI) terms obtained with both virtual and real photons, we are able to reproduce the results reported by Decker and Finkemeier~\cite{DF}:
\begin{equation} \label{SI_total}
\delta R_{\tau/P}\big|_{\mathrm{SI}}=\frac{\alpha}{2\pi}\left\{\frac{3}{2} \mathrm{log}\frac{M_\tau^2 }{m_P^2}- 3\, \mathrm{log}\frac{m_\mu^2}{m_P^2}+\frac{3}{2}  +g\!\left(\frac{m_P^2}{M_\tau^2}\right)-f\bigg(\frac{m_\mu^2}{m_P^2}\bigg)\right\}\,,
\end{equation}
where $g(x)$ and $f(x)$ are given by
\eq{
g(x)=&2\left(\frac{1\!+\!x}{1\!-\!x}\log x-2\right)\log(1\!-\!x)-\frac{x(2\!-\!5x)}{2(1\!-\!x)^2}\log x+4\frac{1\!+\!x}{1\!-\!x}\mathrm{Li}_2(x)+\frac{x}{1\!-\!x}\left(\frac{3}{2}-\frac{4}{3}\pi^2\right),
}
and
\eq{
f(x)=&2\left(\frac{1\!+\!x}{1\!-\!x}\log x-2\right)\log(1\!-\!x)-\frac{x(8\!-\!5x)}{2(1\!-\!x)^2}\log x+4\frac{1\!+\!x}{1\!-\!x}\mathrm{Li}_2(x)-\frac{x}{1\!-\!x}\left(\frac{3}{2}+\frac{4}{3}\pi^2\right),
}
with Li$_2(x)=-\displaystyle\int_0^x \mathrm{d}t \frac{\log(1-t)}{t}\,.$
We agree with the numerical values reported in ref.~\cite{DF}: 
\begin{eqnarray}
\delta R_{\tau/\pi}\big|_{\mathrm{SI}}&=&1.05\%, \nonumber \\
\delta R_{\tau/K}\big|_{\mathrm{SI}}&=&1.67\%.
\end{eqnarray}
From now on, errors are not reported for contributions where the uncertainties are negligible for the level of accuracy of this analysis, that is, lower than 0.01\%.

Note that $G(x)$ and $F(x)$ of (\ref{eq:indratetau}) and (\ref{eq:indrate}) can be related to $g(x)$ and $f(x)$ of (\ref{SI_total}) by \begin{equation} \label{FfGg}
G(x)\,=\,\frac{g(x)}{2}+\frac{19}{8}-\frac{\pi^2}{3} \,, \qquad \quad F(x)\,=\,\frac{3}{2} \log x + \frac{f(x)}{2} +\frac{13}{8} -\frac{\pi^2}{3}\,.
\end{equation}

\subsection*{Structure-dependent contributions}

As far as the  structure-dependent (SD) contributions are concerned, the case with real photons (rSD)  for $P \to\mu \nu_\mu \gamma$ can be extracted from (7.14) and (7.16) of ref.~\cite{CR2}, namely:
\begin{eqnarray}
\delta_{\pi \mu}\big|_{\mathrm{rSD}}=-1.3\cdot 10^{-8},\qquad \quad \delta_{K\mu}\big|_{\mathrm{rSD}}=-1.7\cdot10^{-5}.
\end{eqnarray}
Meanwhile, for $\tau \to P \nu_\tau \gamma$ it can be read from ref.~\cite{Guo:2010dv}:
\begin{eqnarray} \label{rSD_tau}
\delta_{\tau \pi}\big|_{\mathrm{rSD}}=0.15\%, \qquad \quad \delta_{\tau K}\big|_{\mathrm{rSD}}=(0.18\pm 0.05 )\%.
\end{eqnarray}
This gives~\cite{nosotros}
\begin{eqnarray}
\delta R_{\tau/ \pi}\big|_{\mathrm{rSD}}&=&0.15\%,\nonumber\\
\delta R_{\tau/K}\big|_{\mathrm{rSD}}&=&(0.18\pm 0.05)\%.
\end{eqnarray}
being the terms from the $P$ decay negligible.

In the case of $P_{\mu 2}$ the structure-dependent contribution with virtual photons (vSD) can be evaluated using \eqref{eq:indrate} and the numerical values for $c_n^{(P)}$ given in table~\ref{tab:tab1}: 
\begin{eqnarray} \label{vSD_P}
\delta_{\pi \mu}\big|_{\mathrm{vSD}}=(0.54\pm 0.12)\%\,,\qquad \quad \delta_{K\mu}\big|_{\mathrm{vSD}}=(0.43\pm 0.12)\%.
\end{eqnarray}
The completely new calculation that we have had to carry out in our theoretical framework for the analysis of $R_{\tau/P}$ is the vSD contributions for $\tau \to P \nu_\tau$. The relevant Feynman diagram was shown in figure \ref{feynman} and the intermediate steps were given in section~\ref{vSDtau} and appendix~\ref{avsd}. We find 
\begin{eqnarray} \label{vSD_tau}
\delta_{\tau \pi}\big|_{\mathrm{vSD}}=-(0.48\pm 0.56) \%\,,\qquad \quad \delta_{\tau K}\big|_{\mathrm{vSD}}=-(0.45\pm 0.57)\%.  
\end{eqnarray}
Now, using $\delta R_{\tau/P}|_{\text{vSD}}=\delta_{\tau P}|_{\text{vSD}}-\delta_{P\mu}|_{\text{vSD}}$
it is found  that~\cite{nosotros}
\begin{eqnarray}
\delta R_{\tau/ \pi}\big|_{\mathrm{vSD}}&=&-(1.02\pm 0.57 )\%,\nonumber\\
\delta R_{\tau/K}\big|_{\mathrm{vSD}}&=&-(0.88\pm 0.58)\%.
\end{eqnarray}
 
A trustworthy estimation of the uncertainties in (\ref{vSD_tau}) is determinant, since it is the most important source of error in $\delta R_{\tau/P}$. Note the difference between the theoretical framework for the calculation of $\Gamma \left(P\to \mu \nu_{\mu}[\gamma]\right)$ and $\Gamma \left( \tau \to P \nu_{\tau}[\gamma] \right)$:
\begin{enumerate}
\item In $P_{\mu 2}$ decays the computation is performed in the context of Chiral Perturbation Theory (ChPT), the effective field theory of QCD to be used at low energies and, consequently, basically a model-independent calculation. As explained previously, the only model dependence is the determination of the local counterterms and in ref.~\cite{CR1,CR2} it was done by matching ChPT with the effective approach at higher energies, the large-$N_C$ extension including resonances we have quoted previously~\cite{RChT}. From (\ref{eq:indrate}) and the value of $c_1^{(P)}$ given in table~\ref{tab:tab1}, it can be extracted that the uncertainty of this estimation gives an error of approximately $\pm 0.11\%$.
\item In $\tau$ decays, and because of the energy scale at hand, the calculation is done directly with the large-$N_C$ extension of ChPT~\cite{RChT}, so the matching procedure to estimate the unknown counterterms is not possible in this case. The uncertainty of this determination, see (\ref{vSD_tau}), is $\pm 0.56\%$ and $\pm 0.57\%$ for the pion and kaon case, respectively, to be compared to the $\pm 0.11\%$ of $P$ decays. Taking into account this impossibility of following a matching procedure, we consider two sources to estimate the uncertainties of $\delta_{\tau P}\big|_{\mathrm{vSD}}$:
\begin{enumerate}
\item First of all, and in order to assess the model-dependence of the effective approach, we have combined the two scenarios explained in section~\ref{sdff}, see appendix~\ref{avsd}. Although we take as our main framework the more refined analysis of scenario (b), summarized in the tilded form factors of (\ref{scenarioB_vectorpart}) and (\ref{scenarioB_axialpart})~\cite{Guo:2010dv, Guevara:2013wwa}, we have also calculated  $\delta_{\tau P}\big|_{\mathrm{vSD}}$ with scenario (a), summarized in the barred form factors of (\ref{scenarioA})~\cite{Guo:2010dv, Guevara:2013wwa}. We have taken as our first source of error in (\ref{vSD_tau}) one half of the deviation in $\delta_{\tau P}\big|_{\mathrm{vSD}}$ between the two scenarios, resulting in $\pm 0.22\%$ and $\pm 0.24\%$ for the pion and kaon case, respectively.
\item Secondly, and in order to estimate the unknown local counterterms in $\delta_{\tau P}\big|_{\mathrm{vSD}}$, we use their dependences on the renormalization scale, which are known from our calculation, see the divergences of appendix~\ref{avsd}. We have considered as the second source of uncertainty in (\ref{vSD_tau}) one half of the running of the counterterms between $0.5\,$and $1.0\,$GeV, giving $\pm 0.52\%$. A similar procedure was followed in ref.~\cite{Marciano}. It is convenient to stress that we follow a conservative attitude in this estimation: bearing in mind that the first resonances are included in the theoretical framework for $\tau$ decays, their counterterms are expected to be smaller than in $P_{\mu 2}$; notwithstanding, with the effect of the running we account for here, the counterterms affecting $\delta_{P \mu}\big|_{\mathrm{vSD}}$ imply similar corrections to the estimation we consider in $\delta_{\tau P}\big|_{\mathrm{vSD}}$. This can be understood as a subsequent check, that considering further running of the counterterms is not physically motivated.

\end{enumerate}
Adding quadratically these two uncertainties, one gets the errors of (\ref{vSD_tau}): $\pm 0.56\%$ and $\pm 0.57\%$ for the pion and the kaon case, respectively. It is interesting to stress that although the loop integrals with resonances used in the vSD contributions for $P_{\mu 2}$ and $\tau$ decays are related, the consequent correlation between the uncertainties in these decays is negligible, once the uncertainty in $\tau$ decays comes mostly from the estimation of the counterterm and dominates completely the ratio, compare the uncertainties in (\ref{vSD_P}) and (\ref{vSD_tau}).
\end{enumerate}

\begin{table}[t!!!!]
\begin{center}
\begin{tabular}{|c|c|c|c|}
\hline
  Contribution & $\delta R_{\tau/\pi}$   & $\delta R_{\tau/K}$ &  ref.  \\[5pt]
\hline \hline 
SI &  $+1.05\%$& $+1.67\%$ &\cite{DF} \\
rSD  &$+0.15\%$   &$+(0.18\pm 0.05)\%$ & \cite{CR1,CR2,Guo:2010dv} \\
vSD & $-(1.02\pm 0.57 )\%$& $-(0.88\pm 0.58)\%$ & new~\cite{nosotros} \\
\hline \hline
Total & $+(0.18\pm 0.57 )\%$ & $+(0.97\pm 0.58 )\%$ & new~\cite{nosotros}
\\
\hline
\end{tabular}
\end{center}
\caption{Numerical values of the different photonic contributions to $\delta R_{\tau/P}$: Structure Independent (SI), real-photon Structure Dependent (rSD) and virtual-photon Structure Dependent (vSD)~\cite{nosotros}. We do not report uncertainties for contributions where the errors are lower than 0.01\%.}
\label{tab:tab2}
\end{table}

\subsection*{Final result}

We show a summary of the different contributions to $\delta R_{\tau/P}$ in table~\ref{tab:tab2}, which lead our final result~\cite{nosotros}:
\begin{eqnarray} 
\delta R_{\tau/\pi} &=& (0.18\pm 0.57 )\% , \nonumber \\
 \delta R_{\tau/K}&=& (0.97\pm 0.58 )\%. \label{finalresult}
\end{eqnarray} 
As it has been explained previously, the dominant uncertainties come from $\delta_{\tau P}\big|_{\mathrm{vSD}}$.

It is worth comparing our result with ref.~\cite{DF}, $\delta R_{\tau/\pi}=(0.16\pm0.14)\%$ and $\delta R_{\tau/K}=(0.90\pm0.22)\%$. We want to highlight that although their central values agree remarkably, this is solely a coincidence, since the one-sigma confidence intervals overlap only at the $25(38)\%$ level for the $\pi(K)$ case; in our opinion uncertainties were underestimated in ref.~\cite{DF}, having approximately the size expected in a purely Chiral Perturbation Theory computation~\cite{CR1,CR2}, much more model-independent than this calculation, where resonances need to be included taking into account the mass of the $\tau$. Moreover, as it has been remarked in the introduction, the hadronization of the QCD currents of ref.~\cite{DF} is different for real- and virtual-photon corrections, does not satisfy the high-energy behavior dictated by QCD, violates unitarity, analyticity and the chiral limit, and a cutoff is used to regulate the loop integrals, splitting artificially long- and short-distance regimes. A comparison between the results for the different contributions to  $\delta R_{\tau/\pi}$ obtained in the present work and the one of ref.~\cite{DF}  is given in appendix \ref{DF-ours}.

\section{Applications}
\label{appli}

\subsection{Total radiative corrections to $\tau \to P \nu_\tau[\gamma]$ decays}

A remarkable application of our results is to report the total radiative corrections of the individual $\tau \to P \nu_\tau[\gamma]$ decays,
\begin{eqnarray}
\Gamma_{\tau_{P2[\gamma]}}&=& 
\Gamma^{(0)}_{\tau_{P2}}
\,S_{\rm EW}\bigg\{ 1 + \frac{\alpha}{\pi}  \,  G (m_P^2/M_\tau^2)  \bigg\}
\bigg\{  1 - \frac{3\alpha}{2\pi}  \log \frac{m_\rho}{M_\tau} +  \delta_{\tau P}\big|_{\mathrm{rSD}} +  \delta_{\tau P}\big|_{\mathrm{vSD}}\bigg\} \nonumber \\&=& \Gamma^{(0)}_{\tau_{P2}} \,S_{\rm EW}\, \bigg( 1 +\delta_{\tau P} \bigg) \,,
\label{eq:indratetau2}
\end{eqnarray}
where the tree-level result $\Gamma^{(0)}_{\tau_{P2}}$ was shown in (\ref{eq:indratetauLO}) and $S_{\rm EW}=1.0232$ denotes the resumed universal short-distance electroweak corrections~\cite{Marciano}. Taking into account (\ref{FfGg}), it is straightforward to extract the total radiative corrections including all remaining radiative SI and SD corrections $\delta_{\tau P}$ ,
\eq{
\delta_{\tau P} =  \frac{\alpha}{2\pi} \bigg[ g\left(\frac{m_P^2}{M_\tau^2}\right) + \frac{19}{4} -\frac{2\pi^2}{3} - 3\log \frac{m_\rho}{M_\tau} \bigg]+  \delta_{\tau P}\big|_{\mathrm{rSD}} +  \delta_{\tau P}\big|_{\mathrm{vSD}}\,.\label{deltaradiativetau}
}
By using (\ref{rSD_tau}) and (\ref{vSD_tau}), it is found~\cite{nosotros}
\begin{equation}
\delta_{\tau \pi} = -( 0.24 \pm 0.56 ) \%, \qquad \quad
\delta_{\tau K} = -(0.15 \pm 0.57) \%. \label{deltatauP}
\end{equation}

\subsection{Lepton universality test}

In the introduction we stressed that the main motivation in order to determine $R_{\tau / P}$ was to test the lepton universality via (\ref{MainDef}),
\begin{equation}
R_{\tau / P}\equiv \frac{ \Gamma \left( \tau \to P \nu_{\tau}[\gamma] \right)}{\Gamma \left(P\to \mu \nu_{\mu}[\gamma]\right)}= \left\vert \frac{g_\tau}{g_\mu} \right\vert^2 R_{\tau/P}^{0}(1+\delta R_{\tau / P}), \label{MainDef}
\end{equation}
where the leading-order result $R_{\tau/P}^{(0)}$ was given in (\ref{LO}) and $|g_\tau/g_\mu|=1$ according to LU. Considering our results reported in (\ref{finalresult}) and the current experimental data~\cite{Zyla:2020zbs}, we find~\cite{nosotros} 
\begin{align} 
&\left|\frac{g_\tau}{g_\mu}\right|_\pi\!=\!0.9964 \!\pm\! 0.0028_{\mathrm{th}}\!\pm\! 0.0025_{\mathrm{exp}} \!=\! 0.9964\pm 0.0038, \nonumber \\
 &\left|\frac{g_\tau}{g_\mu}\right|_K\!=\!0.9857\!\pm\!  0.0028_{\mathrm{th}}\!\pm\! 0.0072_{\mathrm{exp}} \!=0.9857\pm 0.0078, 
\end{align}
at $0.9\sigma$ and $1.8\sigma$ of lepton universality, respectively. These results should be compared with the HFLAV analysis of ref.~\cite{Amhis:2019ckw}, $\left|g_\tau/g_\mu\right|_\pi=0.9958\pm0.0026$ and $\left|g_\tau/g_\mu\right|_K=0.9879\pm0.0063$, at $1.6\sigma$ and $1.9\sigma$ of LU ($1.4\sigma$ and $2.0\sigma$ in ref.~\cite{Pich:2013lsa},  making use of the PDG input~\cite{Zyla:2020zbs}). Let us stress again that $\delta R_{\tau/P}$ was taken from ref.~\cite{DF} in refs.~\cite{Amhis:2019ckw,Pich:2013lsa}. Note that our results are approaching  LU. 

\subsection{CKM unitarity test}

The CKM unitarity can be tested by considering the ratio
\begin{equation}\label{eq:Vusdet}
\frac{\Gamma(\tau\to K\nu_\tau[\gamma])}{\Gamma(\tau\to \pi\nu_\tau[\gamma])}\,=\,\left|\frac{V_{us}}{V_{ud}}\right|^2 \frac{F_K^2}{F_\pi^2}\frac{(1-m_{K}^2/M_\tau^2)^2}{(1-m_{\pi}^2/M_\tau^2)^2}\left(1+\delta\right).
\end{equation}
From (\ref{eq:indratetau2}) one can determine $\delta$ easily by using our values of (\ref{deltatauP}),
\begin{equation} \label{delta}
\delta = \delta_{\tau K}-  \delta_{\tau P} =  (0.10\pm0.80)\%\,.
\end{equation}
Taking masses and branching ratios from the PDG~\cite{Zyla:2020zbs} and meson decay constants from the FLAG analysis~\cite{Aoki:2019cca}, $F_K/F_\pi=1.1932\pm 0.0019$, one gets~\cite{nosotros}
\begin{equation}  \label{ourVusVud}
 \bigg|\frac{V_{us}}{V_{ud}}\bigg|\!=\!0.2288 \!\pm\! 0.0010_{\mathrm{th}} \!\pm\! 0.0017_{\mathrm{exp}} \!=\! 0.2288\pm0.0020,
\end{equation}
which is $2.1\sigma$ away from unitarity if we consider $|V_{ud}|=0.97373\pm0.00031$ from ref.~\cite{Hardy:2020qwl}. It is convenient to stress that again we have taken a conservative attitude in the estimation of the errors in (\ref{delta}), for we have directly propagated those of (\ref{deltatauP}). Alternatively, by recalculating directly the uncertainties of the difference $\delta_{\tau K}-  \delta_{\tau P}$, that of (\ref{delta}) drops to $\pm 0.05\%$, which would imply $\pm 0.0004_{\mathrm{th}}$ and $\pm 0.0018$ in (\ref{ourVusVud}). The experimental error dominates, so the change is negligible and $|V_{us}/V_{ud}|$ moves from $2.1\sigma$ to $2.2\sigma$ away from unitarity, an insignificant shift. Our result of (\ref{ourVusVud}) is compatible with the value reported recently in ref.~\cite{Seng:2021nar}, $|V_{us}/V_{ud}|=0.2291\pm0.0009$, obtained in the context of kaon semileptonic decays; note that our error does not reach this level of uncertainty due to the lack of statistics in $\tau$ decays, being this an important motivation to improve the related measurements in Belle II~\cite{Belle-II:2018jsg}.

Instead one can extract $|V_{us}|$ directly from the $\tau \to K \nu_\tau[\gamma]$ decay,
\begin{eqnarray}
\Gamma_{\tau_{K2[\gamma]}}&=& 
 \Gamma^{(0)}_{\tau_{K2}} \, S_{\rm EW} \, \bigg( 1 +\delta_{\tau K} \bigg) \,,
\label{eq:indratetau3}
\end{eqnarray}
being $\Gamma^{(0)}_{\tau_{P2}}$ the tree-level result, which was given in (\ref{eq:indratetauLO}). Considering masses and branching ratios from the PDG~\cite{Zyla:2020zbs}, meson decay constants from the FLAG analysis~\cite{Aoki:2019cca}, $\sqrt{2} F_K=(155.7\pm0.3)~$MeV, $S_{\rm EW}=1.0232$ from ref.~\cite{Marciano} and our value of $\delta_{\tau K}$ in (\ref{deltatauP}), it is found~\cite{nosotros}
\begin{equation} \label{ourVus}
 |V_{us}|=0.2220 \pm 0.0008_{\mathrm{th}} \pm 0.0016_{\mathrm{exp}} = 0.2220\pm 0.0018, 
\end{equation}
at $2.6\sigma$ from unitarity by considering again $|V_{ud}|=0.97373\pm0.00031$ from ref.~\cite{Hardy:2020qwl}. We can compare these numbers with the result reported in ref.~\cite{Amhis:2019ckw}, $ |V_{us}|=0.2234 \pm 0.0005_{\mathrm{th}} \pm 0.0014_{\mathrm{exp}} = 0.2234\pm 0.0015$, at $1.3\sigma$ from unitarity. The different experimental inputs are mostly responsible for these  disagreement. Note that again (\ref{ourVus}) is compatible with recent ref.~\cite{Seng:2021nar}, $|V_{us}|=0.2231\pm0.0006$, obtained by considering kaon semileptonic decays.

\subsection{Probing non-standard interactions}

One can also use our results to search for non-standard interactions in $\tau \to P \nu_\tau[\gamma]$ decays, 
\begin{equation}
\Gamma(\tau\to P\nu_\tau[\gamma])=  \Gamma^{(0)}_{\tau_{P2}}  \left|\frac{\widetilde{V}_{uD}}{V_{uD}}\right|^2\!\! S_{\rm EW}\! \left( 1 +\delta_{\tau P}  + 2 \Delta^{\tau P}\right) \,,
\end{equation}
being $\Gamma^{(0)}_{\tau_{P2}}$ the tree-level result, given in (\ref{eq:indratetauLO}), and $D=d,s$ for $P=\pi,K$, respectively.  $\Delta^{\tau P}$ includes the tree-level new-physics corrections not absorbed in $\widetilde{V}_{uD}=(1 + \epsilon^e_L + \epsilon^e_R )V_{uD}$~\cite{EFTtaudecays1,EFTtaudecays2}, directly incorporated by taking 
$V_{uD}$ from nuclear $\beta$ decays,
\begin{equation}
\Delta^{\tau P} = \epsilon^\tau_L-\epsilon^e_L-\epsilon^\tau_R-\epsilon^e_R-\frac{m_P^2}{M_\tau(m_u+m_D)}\epsilon^\tau_P \,.
\end{equation}
Considering again $|V_{ud}|=0.97373\pm0.00031$ from ref.~\cite{Hardy:2020qwl}, masses and branching ratios from the PDG~\cite{Zyla:2020zbs}, meson decays constants from the FLAG analysis~\cite{Aoki:2019cca}, $\sqrt{2} F_\pi=(130.2\pm0.8)~$MeV and $\sqrt{2} F_K=(155.7\pm0.3)~$MeV, $S_{\rm EW}=1.0232$ from ref.~\cite{Marciano} and our values of $\delta_{\tau P}$ and $|V_{us}/V_{ud}|$ in (\ref{deltatauP}) and (\ref{ourVusVud}), respectively, one finds~\cite{nosotros}
 \begin{equation}
\Delta^{\tau \pi} = -(0.15\pm0.72)\%, \qquad \quad
\Delta^{\tau K} =-(0.36\pm1.18)\%\,,
\end{equation}
which update the results reported in ref.~\cite{EFTtaudecays1}, $\Delta^{\tau \pi} = -(0.15 \pm 0.67 ) \%$, and in ref.~\cite{EFTtaudecays2}, $\Delta^{\tau \pi} = -(0.12 \pm 0.68 ) \%$ and $\Delta^{\tau K} = -(0.41 \pm 0.93) \%$. Note that all these numbers are consistent with each other and compatible with zero. The values have been given in the $\overline{\mathrm{MS}}$-scheme and at a scale of $\mu=2$ GeV. 

\section{Conclusions}
\label{Conclu}

There were both phenomenological and theoretical reasons to address the study of $\delta R_{\tau/P}$ ($P=\pi,K$). First of all, we are facing a very convenient observable to test the lepton universality and in the literature there are some disagreements in $|g_\tau / g_\mu |$ depending on the process at hand. Secondly, we have highlighted some inconsistencies in the analysis performed in ref.~\cite{DF}. Although our result, $\delta R_{\tau/\pi}=(0.18\pm 0.57 )\%$  and $\delta R_{\tau/K}=(0.97\pm 0.58 )\%$, is consistent with the result reported more than twenty-five years ago in ref.~\cite{DF}, our approach has much more robust assumptions, resulting in a reliable uncertainty. As a by-product of our main analysis, we report the theoretical radiative corrections in $\tau \to P \nu_{\tau}[\gamma]$ decays: $\delta_{\tau \pi} = -( 0.24 \pm 0.56 ) \%$ and $\delta_{\tau K} = -(0.15 \pm 0.57) \%$.

We have applied our results to different themes:
\begin{enumerate}
\item Lepton universality test. Our values of $\delta R_{\tau/P}$ imply $\left|g_\tau/g_\mu\right|_\pi=0.9964\pm 0.0038$ and $\left|g_\tau/g_\mu\right|_K=0.9857\pm 0.0078$ (at $0.9\sigma$ and $1.8\sigma$ of lepton universality), to be compared to the results of the HFLAV analysis of ref.~\cite{Amhis:2019ckw}, $\left|g_\tau/g_\mu\right|_\pi=0.9958\pm0.0026$ and $\left|g_\tau/g_\mu\right|_K=0.9879\pm0.0063$ (at $1.6\sigma$ and $1.9\sigma$ of lepton universality), obtained by using the value of $\delta R_{\tau/P}$ given in ref.~\cite{DF}.
\item CKM unitarity test. By considering the $\tau$ decays studied in this work, we have extracted the CKM elements $|V_{us}/V_{ud}|= 0.2288\pm0.0020$ and  $|V_{us}|= 0.2220\pm 0.0018$, at $2.1\sigma$ and $1.5\sigma$ from unitarity, respectively. 
\item Searching for non-standard interactions. We have constrained the new-physics corrections to $\tau \to P \nu_\tau[\gamma]$ decays, $\Delta^{\tau \pi} = -(0.15\pm0.72)\cdot10^{-2}$ and $\Delta^{\tau K} =-(0.36\pm1.18)\cdot10^{-2}$, compatible with the values reported in refs.~\cite{EFTtaudecays1,EFTtaudecays2}.
\end{enumerate}

\section*{Acknowledgments}

We wish to thank V. Cirigliano, M. Gonz\'alez-Alonso and A. Pich for their helpful comments. This work has been supported in part by the Universidad Cardenal Herrera-CEU [INDI20/13]; by the Generalitat Valenciana [PROMETEU/2021/071)] and by the Spanish Government [MCIN/AEI/10.13039/501100011033, Grant No. PID2020-114473GB-I00]. M.A.A.U. is funded by Conacyt through the `Estancias posdoctorales nacionales' program. G.L.C. was supported by Ciencia de Frontera Conacyt project No. 428218. G.H.T. and P.R. acknowledge the support of C\'atedras Marcos Moshinsky (Fundaci\'on Marcos Moshinsky). The work of G.H.T. is supported by the postdoctoral fellowship program DGAPA-UNAM and DGAPA-PAPIIT UNAM [Grant No. IN110622].

\appendix

\section{One-loop functions for vSD contributions to $\tau\to P\nu_{\tau}$ decay }\label{avsd}

The Feynman rules necessary to describe the effective structure-dependent vector and axial currents, represented by the gray shaded box in figure \ref{feynman}, were obtained directly from eqs. (\ref{eq:V}) and (\ref{eq:A}), respectively.

For completeness, we report here the definition and decomposition of Passarino-Veltman functions \cite{Ellis:2011cr} appearing in the computation of the virtual-photon structure-dependent (vSD) corrections to $\tau(p_\tau)\to P(p)\nu_\tau(q)$ described in the
main text. For $D=4-2\epsilon$, and introducing $p_0=0$ and $p_{ij}^2=(p_i-p_j)^2$, we have the following definitions:

$\bullet$ Two-point functions
\eq{
\frac{i}{16\pi^2}\big\{B_0, B_\mu, B_{\mu\nu} \big\}({\rm args})_2&=\mu^{2\epsilon}\int \frac{d^Dq}{(2\pi)^D}\big\{1,q_\mu,q_\mu q_\nu \big\}\prod_{i=0}^1\frac{1}{[(q+p_i)^2-m_i^2]},}
where $({\rm args})_2=(p_{10}^2,m_0^2,m_1^2)$. 

$\bullet$ Three-point functions
\eq{
\frac{i}{16\pi^2}\big\{C_0, C_\mu, C_{\mu\nu}, C_{\mu\nu\rho} \big\}({\rm args})_3&=\mu^{2\epsilon}\int \frac{d^Dq}{(2\pi)^D}\big\{1,q_\mu,q_\mu q_\nu, q_\mu q_\nu q_\rho \big\}\prod_{i=0}^2\frac{1}{[(q+p_i)^2-m_i^2]},
}where $({\rm args})_3=(p_{10}^2,p_{12}^2,p_{20}^2,m_0^2,m_1^2,m_2^2)$. 

$\bullet$ Four-point functions
\eq{
\frac{i}{16\pi^2}\big\{D_0, D_\mu, D_{\mu\nu}, D_{\mu\nu\rho}, D_{\mu\nu\rho\sigma} \big\}({\rm args})_4&=\mu^{2\epsilon}\int \frac{d^Dq}{(2\pi)^D}\big\{1,q_\mu,q_\mu q_\nu, q_\mu q_\nu q_\rho, q_\mu q_\nu q_\rho q_\sigma \big\}\nonumber\\
&\quad\quad\quad\times\prod_{i=0}^3\frac{1}{[(q+p_i)^2-m_i^2]},
}
where $
({\rm args})_4=(p_{10}^2,p_{12}^2,p_{23}^2,p_{30}^2,p_{20}^2,p_{13}^2,m_0^2,m_1^2,m_2^2, m_3^2).$\\

The above general fuctions can be decomposed as follows 
\eq{
B_\mu &=p_{1_{\mu}}B_1,\nonumber\\
B_{\mu\nu}&=g_{\mu\nu}B_{00}+p_{1_{\mu}}p_{1_{\nu}}B_{11}.
}
\eq{
C_\mu &=p_{1_{\mu}}C_1+p_{2_{\mu}}C_2,\nonumber\\
C_{\mu\nu}&=g_{\mu\nu}C_{00}+\sum_{i,j=1}^2p_{i_{\mu}}p_{j_{\nu}}C_{ij},\nonumber\\ C_{\mu\nu\rho}&=\sum_{i=1}^2\lambda_{i_{\mu\nu\rho}}C_{00i}+\sum_{i,j,k=1}^2 p_{i_{\mu}}p_{j_{\nu}}p_{k_{\rho}}C_{ijk}.
}
\eq{
D_\mu &=\sum_{i=1}^3 p_{i_\mu}D_i,\nonumber\\
D_{\mu\nu} &=g_{\mu\nu}D_{00}+\sum_{i,j=1}^3p_{i_{\mu}}p_{j_{\nu}}D_{ij},\nonumber\\
D_{\mu\nu\rho}&=\sum_{i=1}^3 \lambda_{i_{\mu\nu\rho}}D_{00i}+\sum_{i,j,k=1}^3 p_{i_{\mu}}p_{j_{\nu}}p_{k_{\rho}}D_{ijk},\nonumber\\
D_{\mu\nu\rho\sigma}&=\lambda_{\mu\nu\rho\sigma}D_{0000}+\sum_{i,j=1}^3 \lambda_{{ij}_{\mu\nu\rho\sigma}}D_{00ij}+\sum_{i,j,k,l=1}^3p_{\mu_i}p_{\nu_j}p_{\rho_k}p_{\sigma_l}D_{ijkl}.
} with the definitions $\lambda_{\mu\nu\rho\sigma}=g_{\mu\nu}g_{\rho\sigma}+g_{\mu\rho}g_{\nu\sigma}+g_{\mu\sigma}g_{\nu\rho}$, $\lambda_{i_{\mu\nu\rho}}=g_{\mu\nu}p_{i_{\rho}}+g_{\nu\rho}p_{i_{\mu}}+g_{\rho\mu}p_{i_{\nu}}$, $\lambda_{{ij}_{\mu\nu\rho\sigma}}=g_{\mu\nu}p_{i_{\rho}}p_{j_{\sigma}}+g_{\nu\rho}p_{i_{\mu}}p_{j_{\sigma}}+g_{\mu\rho}p_{i_{\nu}}p_{j_{\sigma}}+g_{\mu\sigma}p_{i_{\nu}}p_{j_{\rho}}+g_{\nu\sigma}p_{i_{\mu}}p_{j_{\rho}}+g_{\rho\sigma}p_{i_{\mu}}p_{j_{\nu}}$.

We have cross-checked our results using both \textit{Feyncalc} \cite{Shtabovenko:2020gxv} and \textit{Package-X} \cite{Patel:2016fam}. Here, we reported the shortest, among equivalent, expressions. Analytical expressions for the scalar and the Passarino-Veltman functions can be obtained in principle directly with \textit{Package-X}. Nevertheless, for compactness, we preferred to evaluate them directly using the package Collier \cite{Denner:2016kdg}. 
We introduce the following notation to distinguish functions with different arguments appearing in our results. For two-point functions ($a,b=0,1$)
\eq{
B_{\{a, ab\}}({\rm args})_2&\equiv B_{\{a, ab\}}(M_\tau^2,0,M_\tau^2),\nonumber\\
\tilde{B}_{\{a, ab\}}({\rm args})_2&\equiv B_{\{a, ab\}}(0,M_A^2,M_\tau^2), \nonumber\\
B^\star_{\{a, ab\}}({\rm args})_2 &\equiv B_{\{a, ab\}}(M_\tau^2,M_V^2,M_\tau^2).
\label{BsAppx}} For three-point functions ($a,b,c=0,1,2$)
\eq{
\bar{C}_{\{a,ab,abc\}}({\rm args})_3&=C_{\{a,ab,abc\}}(m_P^2,M_\tau^2,0,m_P^2,M_V^2,M_\tau^2),\nonumber\\
\tilde{C}_{\{a,ab,abc\}}({\rm args})_3&=C_{\{a,ab,abc\}}(0,M_\tau^2,M_\tau^2,0,M_V^2,M_\tau^2),\nonumber\\
C^\star_{\{a,ab,abc\}}({\rm args})_3&=C_{\{a,ab,abc\}}(M_\tau^2,0,m_P^2,0,M_\tau^2,M_A^2).
\label{CsAppx}}For four-point functions ($a,b,c,d,e=0,1,2,3$)
\eq{
D_{\{a, ab,abc,abcd,abcde\}}({\rm args})_4&=D_{\{a, ab,abc,abcd,abcde\}}(M_\tau^2,M_\tau^2,m_P^2,m_P^2,0,0,0,M_\tau^2,M_V^2,M_V^2),\nonumber\\
\bar{D}_{\{a, ab,abc,abcd,abcde\}}({\rm args})_4&=D_{\{a, ab,abc,abcd,abcde\}}(M_\tau^2,M_\tau^2,m_P^2,m_P^2,0,0,0,M_\tau^2,M_V^2,M_A^2).
\label{DsAppx}}
Considering the above definitions, we have that:
\begin{itemize} 
\item  The relevant one-loop contributions for \textbf{\textit{scenario (a)}}, obtained by considering the form factors of (\ref{Bformfactor}) and (\ref{scenarioA}), read
\eq{
f^{(a)}_V&=\frac{-N_c}{24\pi^2F_P} \bigg\{\left(m_P^2-M_\tau^2\right)\left(B_{11}+\frac{1}{6}\right)-6B_{00}+\frac{5}{3}M_\tau^2 \bigg\}, \nonumber \\
f_{A}^{(a)}&=\frac{F_A^2}{2F_P}\bigg\{2(M_A^2-m_P^2)C^\star_0-2(M_\tau^2+M_A^2)(C^\star_1+C^\star_2)+3{B_0}+2\tilde{B}_0-4 \bigg\}, \nonumber\\
f_{B}&=2F_P\bigg\{ B_0^\star-M_\tau^2\left[\bar{C}_1+2(\tilde{C}_{22}-\tilde{C}_{222})\right]+m_P^2\bar{C}_0+4\left(3\tilde{C}_{002}-2\tilde{C}_{00}\right)+\frac{7}{3}\bigg\},
}
where we follow the notation of section~\ref{vSDtau} and we use the superscript $(a)$ for {\it scenario (a)}. The divergent parts of the above expressions are
\eq{
f_{V\textrm{div}}^{(a)}= \frac{-N_c}{24\pi^2F_P}\left(\frac{m_P^2-4M_\tau^2}{3}\right)\Delta,\quad f_{A\textrm{div}}^{(a)}=\frac{5F_A^2}{2F_P}{\Delta},\quad f_{B\textrm{div}}&=-{4}F_P{\Delta},
}where $\Delta=2\mu^{D-4}/(4-D) -\gamma_{\mathrm{Euler}}+\log 4\pi$.

Note that the counterterm appearing in $\delta_{\tau P}\big|_{\mathrm{vSD}}$ needs to cancel these divergences, so if one is dealing with \textit{scenario (a)}, one has to consider
\eq{
\frac{\alpha}{4\pi}G_F V_{uD} M_\tau\, \bar{u}(q)\left( 1 +\gamma_5 \right)
u(p_\tau) \left( f_{V\textrm{div}}^{(a)}+f_{A\textrm{div}}^{(a)}+f_{B\textrm{div}} \right).
} 

\item For \textbf{\textit{scenario (b)}}, taking the form factors of (\ref{Bformfactor}), (\ref{scenarioB_vectorpart}) and (\ref{scenarioB_axialpart}), we have that:
\eq{
f_{V}^{(b)}&=\frac{-N_c M_V^4}{24\pi^2 F_P}\left[(m_P^2-M_\tau^2)(D_{13}+D_{11})-6D_{00} \right],\nonumber\\
f_{A}^{(b)}&= f_{A1}^{(b)}+f_{A2}^{(b)}+f_{A3}^{(b)}, \nonumber \\
f_{A1}^{(b)}&=\frac{F_P}{2}\left[4M_\tau^2 \tilde{C}_{222}+2m_P^2 \tilde{C}_{22}+24\tilde{C}_{002}-4\tilde{C}_{00}+\frac{8}{3}\right],\label{fA1}\nonumber\\
f_{A2}^{(b)}&=2\frac{F_P^2}{F_A^2}f_{A_1}^{(a)},\nonumber\\
f_{A3}^{(b)}&=\frac{F_P}{2}\bigg[M_\tau^4\bigg(-4\bar{D}_{1111}+8\bar{D}_{11111}+24\bar{D}_{11113}-8\bar{D}_{1113}+24\bar{D}_{11133}+4\bar{D}_{113}-4\bar{D}_{1133}\nonumber\\
&+8\bar{D}_{11333}-2\bar{D}_{13}+4\bar{D}_{133}\bigg)+m_P^4\bigg(-3\bar{D}_{11}-2\bar{D}_{111}+8\bar{D}_{11133}-14\bar{D}_{113}-4\bar{D}_{1133}\nonumber\\
&+24\bar{D}_{11333}-4\bar{D}_{13}-26\bar{D}_{133}-8\bar{D}_{1333}+24\bar{D}_{13333}+3\bar{D}_3+\bar{D}_{33}-14\bar{D}_{333}-4\bar{D}_{3333}\nonumber\\
&+8\bar{D}_{33333}\bigg)+M_\tau^2 m_P^2\bigg(-8\bar{D}_{111}+16\bar{D}_{11113}-8\bar{D}_{1113}+48\bar{D}_{11133}-20\bar{D}_{113}-16\bar{D}_{1133}\nonumber\\
&+48\bar{D}_{11333}-8\bar{D}_{133}-8\bar{D}_{1333}+16\bar{D}_{13333}-3\bar{D}_{3}-4\bar{D}_{33}+4\bar{D}_{333}\bigg)+4M_\tau^2\bigg(-\bar{D}_{00}\nonumber\\
&+\bar{D}_{001}-12\bar{D}_{0011}+32\bar{D}_{00111}+64\bar{D}_{00113}-12\bar{D}_{0013}+32\bar{D}_{00133}+6\bar{D}_{003}\bigg)\nonumber\\
&+2m_P^2\bigg(5\bar{D}_{00}-20\bar{D}_{001}+64\bar{D}_{00113}-24\bar{D}_{0013}+128\bar{D}_{00133}-30\bar{D}_{003}-24\bar{D}_{0033}\nonumber\\
&+64\bar{D}_{00333}\bigg)-96\bar{D}_{0000}+384D_{00001}+384D_{00003}+16\Bigg],
}
where the functions $f_{A1}^{(b)}$, $f_{A2}^{(b)}$, and $f_{A3}^{(b)}$ come from the the first, second, and third term of the axial form factor defined in (\ref{scenarioB_axialpart}), respectively. Again, we use the superscript $(b)$ for \textit{scenario (b)}. Note that $f_{B}$ does not change comparing to \textit{scenario (a)}, and this is the reason why we do not repeat it. 
Note that $f_{V}^{(b)}$ is free of UV divergences, whereas the UV divergent parts of the $f_{A1}^{(b)}$, $f_{A2}^{(b)}$, and $f_{A3}^{(b)}$ functions are
\eq{
f_{A1\textrm{div}}^{(b)}=-\frac{3F_P}{2}\Delta,\quad
f_{A2\textrm{div}}^{(b)}={5 F_P}{\Delta},\quad
f_{A3\textrm{div}}^{(b)}=-{6 F_P}{\Delta}.
}
The counterterm appearing in $\delta_{\tau P}\big|_{\mathrm{vSD}}$ needs to cancel these divergences, so if one is dealing with {\it scenario (b)}, one is required to consider the divergences given by
\eq{
\frac{\alpha}{4\pi}G_F V_{uD} M_\tau\, \bar{u}(q)\left( 1 +\gamma_5 \right)
u(p_\tau) \left(f_{A_{1}\textrm{div}}^{(b)}+f_{A_{2}\textrm{div}}^{(b)}+f_{A_{3}\textrm{div}}^{(b)}+f_{B\textrm{div}}\right)\,.
} 
\end{itemize}

\section{Comparison with results of ref.~\cite{DF}}\label{DF-ours}

Previous calculations of $\delta R_{\tau / P}$ were given in ref.~\cite{DF}. Following a prescription introduced by Sirlin \cite{Sirlin:1972cs}, these authors have used a modified photon propagator
\eq{\frac{1}{k^2-\lambda^2} =\underbrace{\frac{1}{k^2-\lambda^2}\cdot \frac{\mu^2_{\rm cut}}{\mu^2_{\rm cut}-k^2}}_{\rm long-distance}+\underbrace{\frac{1}{k^2-\mu_{\rm cut}^2}}_{\rm short-distance} \label{php}}
to separate the long- and short-distance photonic contributions. The structure-independent (SI) and structure-dependent (SD) virtual corrections were computed using the first term of eq.~(\ref{php}) in terms of hadronic degrees of freedom, while the short-distance corrections were computed using the second term. As a result, virtual corrections to $R_{\tau/P}$ become $\mu_{\rm cut}$-dependent \cite{DF}. 

In table \ref{comparison} we compare the different contributions to $\delta R_{\tau / \pi}$ obtained from table 3 of ref.~\cite{DF}  and our calculation.\footnote{Note that with the original relative sign between IB and SD contributions in ref.~\cite{Decker:1993ut} (see also our footnote~\ref{footnote1}), their rSD correction reads $+0.17\%$ (see table 2 in ref.~\cite{Decker:1993ut}), much closer to our value of $+0.15\%$.} For the SI contribution in ref.~\cite{DF} we have added to the result obtained in the point-like approximation ($1.05\%$) the term coming from cutting off the loops at $\mu_{\rm{cut}}$ ($-0.21\%$), also reported in table 3 of ref.~\cite{DF}. Results corresponding to $\delta R_{\tau/K}$ are not provided in ref.~\cite{DF} which prevents a comparison.

\begin{table}[t!!!!]
\begin{center}
\begin{tabular}{|c|c|c|}
\hline
  Contribution & ref.~\cite{DF} [$\mu_{\rm cut}=1.5~$GeV]& Present calculation\\
\hline \hline 
SI & $+0.84\%^*$ & $+1.05\%$  \\
rSD & $+0.05\%$ & $+0.15\%$ \\
vSD & $-0.49\%^*$ & $-(1.02\pm 0.57)\%$ \\
short-distance & $-0.25\%^*$ & 0 \\
\hline \hline
Total & $+(0.16\pm 0.14)\%^*$ & $+(0.18\pm 0.57)\%$ \\
\hline
\end{tabular}
\end{center}
\caption{Comparison of photonic contributions to $\delta R_{\tau/\pi}$. Figures with an asterisk are  cutoff-dependent, with $\mu_{\rm cut}=1.5$ GeV \cite{DF}. The quoted error in the radiative correction of ref.~\cite{DF} arises from uncertainties in hadronic parameters of SD contributions and from variations in the cutoff parameter, $\mu_{\rm cut}$.} \label{comparison}
\label{tab:tab2}
\end{table}

We observe that, although central values for the sum of different corrections agree remarkably, this is a mere coincidence. In particular, central values for the structure-dependent corrections are largely different within both approaches.



\end{document}